\documentstyle[aps,preprint,epsf]{revtex}

\newcommand{\ds}{\displaystyle}
\newcommand{\ltwid}{\raise.3ex\hbox{$<$\kern-.75em\lower1ex\hbox{$\sim$}}}
\newcommand{\rtwid}{\raise.3ex\hbox{$>$\kern-.75em\lower1ex\hbox{$\sim$}}}
\renewcommand{\theequation}{\arabic{section}.\arabic{equation}}

\draft

\title{Phase Fluctuations and Single Fermion Spectral Density  in
2D Systems with Attraction}

\author{V.P.~Gusynin\thanks{E-mail: vgusynin@bitp.kiev.ua},
        V.M.~Loktev\thanks{E-mail: vloktev@bitp.kiev.ua},}

\address{Bogolyubov Institute for
         Theoretical Physics,
         03143 Kiev, Ukraine}

\author{and \\ S.G.~Sharapov\thanks{
On leave of absence from Bogolyubov Institute for Theoretical Physics
of the National Academy of Sciences of Ukraine,
03143 Kiev, Ukraine}}

\address{Department of Physics,
         University of Pretoria,
         0002 Pretoria, South Africa}

\date{July 29, 1999}

\begin{document}
\tighten

\maketitle

\begin{abstract}
The effect of static fluctuations in the phase of the order
parameter on the normal and superconducting properties of a 2D
system with attractive four-fermion interaction is studied.
Analytic expressions for the fermion Green's function, its spectral
density, and the density of states are derived in the approximation
where the coupling between the spin and charge degrees of freedom
is neglected.  The resulting single-particle Green's function clearly
demonstrates a non-Fermi liquid behavior. The results show that as the
temperature increases through the 2D critical temperature, the width
of the quasiparticle peaks broadens significantly.
\end{abstract}

General properties: correlation between physical properties
in normal and superconducting states;
74.40.+k
Fluctuations;
74.72.-h
High-$T_c$ compounds \\


\newpage
\section{Introduction}
\setcounter{equation}{0}

One of the most convincing manifestations of the difference between
the BCS scenario and superconductivity in the cuprates is the pseudogap,
or the depletion of a single particle spectral weight around the Fermi
level \cite{Levi}.  This is observed mainly in the underdoped cuprates
where the pseudogap opens in the normal state as the temperature $T$
decreases below the crossover temperature $T^{\ast}$ and extends over a
wide range of $T$.

Due to the complex nature of cuprate systems, there are a number
of theoretical explanations for the pseudogap behavior. One of them
is based on the model of a nearly antiferromagnetic Fermi liquid
\cite{Pines.review}.  Another possible explanation relates the
pseudogap to spin- and/or charge-density waves \cite{Klemm}. A
third direction, which we take in this paper, argues that
precursor superconducting fluctuations may be responsible for the
pseudogap phenomena. Indeed an incoherent pair tunneling experiment
\cite{Scalapino} proposed recently may allow one to answer whether
the superconducting fluctuations are really responsible for the
pseudogap behavior. Furthermore, one cannot exclude the
possibility that the pseudogap is the result of a combination of
various mechanisms, e.g., both spin and superconducting fluctuations.

The precursor superconducting fluctuations have recently been
extensively studied using different approaches. In most cases, the
attractive 2D or 3D Hubbard model was considered. In particular
this model has been studied, both analytically
\cite{Haussmann.1993,Tchernyshyov,Kagan} and numerically
\cite{Serene,Micnas,Haussmann.1994,Gooding}, in the conserving
$T$-matrix approximation that is  ``$\Phi$ derivable'' in the
sense of Baym \cite{Baym}. The non ``$\Phi$ derivable''
$T$-matrix approximation was considered in \cite{Levin}.
In this approach, the pseudogap is related to the resonant pair
scattering of correlated electrons above $T_c$.
For the d-wave pairing, the pseudogap was also studied in
\cite{Nazarenko} (for a review, see \cite{Randeria}) and Monte
Carlo simulations for the 2D attractive Hubbard model were
performed in \cite{Singer}.

It is known, however, that while the $T$-matrix approximation
provides an adequate description of 3D systems at all temperatures,
including the superconducting state with a long-range
order, it fails (see, for example, \cite{Micnas}) to describe the
Berezinskii-Kosterlitz-Thouless (BKT) transition into the state
with an algebraic order, which is only possible in 2D systems. This
is why, in most of the papers cited above, the $T$-matrix
approximation was used to study either 3D systems
\cite{Haussmann.1993,Tchernyshyov,Haussmann.1994,Levin} or 2D
systems above $T_c$ \cite{Serene,Micnas,Nazarenko,Randeria}
in order to avoid the BKT transition, even though it is generally
accepted that 2D models are more relevant for the description of
cuprates \cite{Choy}.

Of course, the superconducting transition itself is not of the BKT
type, because even a weak interplanar coupling produces a transition
in the $d = 3$ XY universality class, sufficiently close to the
transition temperature.
Outside the transition region, however, the low-energy
physics is governed by vortex fluctuations \cite{Sudbo},
and one can expect the 2D model to be especially
relevant for the description of the pseudogap phase. This was
confirmed for the quasi-2D model \cite{Preosti} (see also
\cite{quasi2D}).

Regarding the pseudogap, it is sufficient to consider the case where
$T > T_{c}$ . However, one definitely needs an  approach different from
the $T$-matrix if one wants to study the 2D theory for the entire
temperature range and wants to connect the pseudogap to
the superconducting gap.  An alternative approach overcoming
the above difficulty was proposed in
\cite{Gusynin.JETPLett,Gusynin.JETP,Gusynin.JETPLett1999}. For a
2D system, one should rewrite the complex order field $\Phi(x)$
in terms of its modulus $\rho(x)$ and its phase $\theta(x)$ as
$\Phi(x)=\rho(x)\exp[i \theta(x)]$, which was originally suggested
by Witten in the context of 2D quantum field theory \cite{Witten}.
It is impossible to obtain $\Phi \equiv\langle \Phi(x) \rangle \ne
0$ at finite $T$ because this would correspond to the formation of
symmetry breaking homogeneous long-range order, which is forbidden
by the Coleman---Mermin---Wagner---Hohenberg (CMWH) theorem
\cite{Coleman}.  However, it is possible to obtain $\rho\equiv
\langle\rho(x) \rangle \ne 0$ with
$\Phi = \rho \langle \exp[i\theta(x)] \rangle = 0$ at
the same time because of random fluctuations of the phase
$\theta(x)$ (i.e., because of transverse fluctuations of
the order field originating in the modulus conservation principle
\cite{Patashinskii}). We stress that $\rho\ne 0$ does not imply any
long-range superconducting order (which is destroyed by phase
fluctuations) and, therefore, does not contradict the
abovementioned theorem.

For the simple model studied in
\cite{Gusynin.JETPLett,Gusynin.JETP}, there are
three regions in the 2D phase diagram. The first one is the
superconducting (here, BKT) phase with $\rho \ne 0$ at $T < T_{\rm BKT}$,
where $T_{\rm BKT}$ is the BKT transition temperature, which plays
the role of $T_c$ in pure 2D superconducting systems. In this region,
there is an algebraic order or a power law decay of the
$\langle\Phi^{\ast}\Phi\rangle$ correlations.  The second
region corresponds to the so-called pseudogap phase
($T_{\rm BKT} < T < T_{\rho} $), where $T_\rho$ is the temperature at
which $\rho$ is supposed to become zero. In this phase, $\rho$ is
still non-zero, but the above correlations decay exponentially.
The third is the normal (Fermi-liquid) phase at $T > T_{\rho}$, where
$\rho= 0$. Note that $\Phi$ and all the symmetry violating correlators
like $\langle\Phi(x)\Phi(0)\rangle$ vanish everywhere.

The proposed description of the phase fluctuations and the BKT
transition is very similar to that given by Emery and Kivelson
\cite{Emery}. However, the field $\rho(x)$ does not appear
the phenomenological approach of \cite{Emery}, while in the present
microscopic approach it occurs naturally. We also mention here the
application of similar ideas to the 3D case \cite{Babaev}, where
instead of the 2D temperature $T_{\rm BKT}$ one has
the phase transition temperature in the 3D XY-model, $T_c^{XY}$.

The main quantity of interest in the present paper is the
one-fermion Green's function and the associated spectral function
$A(\omega,{\bf k})=-(1/\pi){\rm Im}G(\omega+i0,{\bf k})$. The
second quantity, being proportional to the intensity of the
angle-resolved photoemission spectrum (ARPES) \cite{Campuzano},
encodes information about the pseudogap and quasiparticles.
Following the approach of
Refs.~\cite{Gusynin.JETPLett,Gusynin.JETP,Gusynin.JETPLett1999}
the Green's function for the charged (physical) fermions is given
by the convolution (in momentum space) of the propagator for
neutral fermions (which has a gap $\rho\neq0$) and the Fourier
transform of the phase correlator
$\langle\exp(i\tau_3\theta(x)/2)\exp(-i\tau_3\theta(0)/2)\rangle$.

Thus, the approximation employed here assumes the absence of
coupling between spin and charge degrees of freedom;  this can
be taken into account at the next stage of approximation.
We demonstrate that the quasiparticle spectral function broadens
considerably when passing from the superconducting to the normal
state, as observed experimentally \cite{Campuzano}. More importantly,
the phase fluctuations result in a non-Fermi liquid behavior of the
system both below and above $T_{\rm BKT}$.

We note that the effect of classical phase fluctuations of
the order field on the spectral properties of underdoped cuprates
has also been analyzed by Franz and Millis \cite{Franz}. Being
experimentally motivated, they could show that the corresponding
photoemission and tunneling data are well accounted for by a simple
model where $d$-wave charge excitations are coupled to supercurrent
fluctuations.

A brief overview of the paper is as follows: In Sec.~\ref{sec:form}, we
present the modulus-phase formalism for the fermion Green's function
and explain why it is so important to use this formalism for the
description of 2D models. In Sec.~\ref{sec:phase.green}, we obtain
and discuss the Green's function of phase fluctuations both below
and above $T_{\rm BKT}$. This expression is then used in
Sec.~\ref{sec:green.fermion} to derive the temperature and retarded
fermion Green's functions. We show that this Green's function exhibits
a non-Fermi liquid behavior. In Sec.~\ref{sec:spectral}, we obtain an
analytic expression for the spectral density of the fermion Green's
function and discuss this result in detail. The density of states
(DOS) is considered in Sec.~\ref{sec:DOS}. Appendix~\ref{sec:A}
contains technical details on the calculation of the long-distance
asymptotic behavior of the phase correlator. Appendix~\ref{sec:B} contains
the derivation of an alternative representation for the fermion
Green's function which is useful in calculating the spectral
density.  The integrals for the DOS  are given in
Appendix~\ref{sec:C}.

\section{The modulus---phase representation for
         the fermion Green's function}
\setcounter{equation}{0}
\label{sec:form}
Our starting point is a continuum version of the two-dimensional
attractive Hubbard model defined by the Hamiltonian density
\cite{Gusynin.JETPLett,Gusynin.JETP,Gusynin.JETPLett1999}
\begin{equation} {\cal H} =
\psi_{\sigma}^{\dagger}(x) \left(- \frac{\nabla^{2}}{2 m} - \mu \right)
    \psi_{\sigma}(x) - V \psi_{\uparrow}^{\dagger}(x)
       \psi_{\downarrow}^{\dagger}(x) \psi_{\downarrow}(x)
       \psi_{\uparrow}(x),
\label{Hamilton}
\end{equation}
where $x= \mbox{\bf r}, \tau$ denotes the space and imaginary time
variables, $\psi_{\sigma}(x)$ is a fermion field with the spin $\sigma
=\uparrow,\downarrow$, $m$ is the effective fermion mass, $\mu$ is
the chemical potential, and $V$ is an effective local attraction
constant; we take $\hbar = k_{B} = 1$. The model with the
Hamiltonian density (\ref{Hamilton}) is equivalent to the model
with an auxiliary BCS-like pairing field, which can be written as
\begin{equation} {\cal H} =
\Psi^{\dagger}(x)\left[\tau_3 \left(- \frac{\nabla^{2}}{2 m} - \mu \right)
 - \tau_{+}\Phi(x) - \tau_{-}\Phi^\ast(x)\right]\Psi(x)+
\frac{|\Phi(x)|^2}{V}
\end{equation}
in terms of Nambu variables
\begin{equation}
\Psi(x) = \left( \begin{array}{c}
\psi_{\uparrow}(x) \\  \psi_{\downarrow}^{\dagger}(x)
\end{array} \right), \qquad
\Psi^{\dagger}(x) = \left( \begin{array}{cc}
\psi_{\uparrow}^{\dagger}(x) \quad \psi_{\downarrow}(x)
\end{array} \right) ,
\label{Nambu.spinors}
\end{equation}
where $\tau_{\pm} = (\tau_{1} \pm i \tau_{2})/2$, $\tau_{3}$ are
the Pauli matrices and
$\Phi(x)= V \Psi^\dagger(x)\tau_{-}\Psi(x)= V \psi_\downarrow \psi_\uparrow$
is the complex order field.

We consider the full fermion Green's function in the
Matsubara finite temperature formalism
\begin{equation}
G(x) = \langle \Psi(x) \Psi^{\dagger}(0) \rangle \,. \label{Green.def}
\end{equation}
For the 3D case of the BCS theory, the frequency-momentum representation
for (\ref{Green.def}) in the mean field approximation is
known to be \cite{Schrieffer}
\begin{equation}
G(i \omega_{n}, \mbox{\bf k}) = - \frac{ i \omega_{n} \hat{I} +
\tau_{3} \xi(\mbox{\bf k}) -
\tau_{+} \Phi - \tau_{-} \Phi^{\ast}}
{\omega_{n}^{2} + \xi^{2}(\mbox{\bf k}) + |\Phi|^{2}},
\label{Green.standard}
\end{equation}
where $\omega_{n} = (2n+1)\pi T$ is the odd (fermion) Matsubara
frequency, $\xi(\mbox{\bf k})$ is the dispersion law of electrons
evaluated from the chemical potential $\mu$, and
$\Phi\equiv\langle\Phi(x)\rangle$ is the complex order parameter.

A problem arises when one tries to apply Eq.
(\ref{Green.standard}) directly to 2D systems, since it has been
proved (see \cite{Coleman}) that nonzero $\Phi$ values are
forbidden. Nevertheless, one can assume that the modulus of the
order parameter $\rho = |\Phi|$ has a nonzero value, while its
phase $\theta(x)$ defined by
\begin{equation}
\Phi(x) = \rho(x) \exp{[ i \theta(x)]}
\label{phase.def}
\end{equation}
is a random quantity. To be consistent with (\ref{phase.def}) one
should also introduce the spin-charge variables for the Nambu
spinors
\begin{equation}
\Psi(x) =\exp[i\tau_3\theta(x)/2]\Upsilon(x)$, \qquad
$\Psi^{\dagger}(x) = \Upsilon^{\dagger}(x)
\exp[-i\tau_3\theta(x)/2],
\label{Nambu.phase}
\end{equation}
where $\Upsilon$ is the neutral fermion field operator. The
strategy of treating charge and spin (neutral) degrees of freedom as
independent seems to be quite useful, and at the same time a very
general feature of 2D systems \cite{Witten,Wiegmann}.

Applying (\ref{Nambu.phase}), we thus split the Green's function
(\ref{Green.def}) into spin and charge parts
\begin{equation}
G_{\alpha \beta}(x) = \sum_{\alpha^\prime, \beta^{\prime}} {\cal
G}_{\alpha^\prime\beta^\prime}(x)
\langle (e^{i\tau_3\theta(x)/2})_{\alpha\alpha^\prime}
(e^{-i\tau_3\theta(0)/2})_{\beta^\prime\beta} \rangle,
\label{Green.splited}
\end{equation}
where
\begin{equation}
{\cal G}_{\alpha\beta}(x)=\langle \Upsilon_{\alpha}(x)
\Upsilon^{\dagger}_{\beta}(0) \rangle
\end{equation}
is the Green's function for neutral fermions. Introducing the
projectors $P_{\pm}={1\over2}(\hat I\pm \tau_3)$ we obtain
\begin{equation}
e^{i\tau_3\theta/2} = P_{+}e^{i\theta/2}+P_{-}e^{-i\theta/2},
\quad
e^{-i\tau_3\theta/2} = P_{-}e^{i\theta/2}+P_{+}e^{-i\theta/2},
\end{equation}
so that (\ref{Green.splited}) can be rewritten as
\begin{equation}
G(x) = \sum_{\alpha,\beta=\pm} P_\alpha{\cal G}(x) P_\beta
\langle \exp [i\alpha\theta(x)/2 ] \exp [- i\beta\theta(0)/2] \rangle,
\label{Green.projectors}
\end{equation}
where $\alpha=\beta$ and $\alpha=-\beta$ correspond to  the diagonal and
non-diagonal parts of the Green's function, respectively.

 For the frequency-momentum representation of
(\ref{Green.projectors}) we have
\begin{equation}
G(i \omega_{n}, \mbox{\bf k}) = T \sum_{m = - \infty}^{\infty} \int
\frac{d^2 p}{(2 \pi)^{2}}
\sum_{\alpha,\beta=\pm}
P_\alpha {\cal G}(i \omega_{m}, \mbox{\bf p}) P_\beta D_{\alpha
\beta} (i \omega_{n} - i \omega_{m}, \mbox{\bf k} - \mbox{\bf p}),
\label{Green.projectors.momentum}
\end{equation}
where
\begin{equation}
{\cal G}(i \omega_{m}, \mbox{\bf k}) =
\int_{0}^{1 / T} d \tau \int d^2 r
\exp [i \omega_{m} \tau - i \mbox{\bf k} \mbox{\bf r}]
{\cal G}(\tau, \mbox{\bf r})
\label{Green.neutral.Fourier}
\end{equation}
 and
\begin{equation}
D_{\alpha \beta}(i \Omega_{n}, \mbox{\bf q}) =
\int_{0}^{1 /T} d \tau \int d^2 r
\exp \left(i \Omega_{n} \tau - i \mbox{\bf q} \mbox{\bf r}\right)
\langle \exp [ i\alpha \theta(\tau, \mbox{\bf r}) / 2 ]
 \exp [ - i\beta \theta(0) / 2 ] \rangle
\label{Green.phase.Fourier}
\end{equation}
is the correlator of phase fluctuations with even (boson)
frequencies $\Omega_{n} = 2 \pi n T$.

There is a good reason to believe (see \cite{Gusynin.JETP}) that
for $T$ close to $T_{\rm BKT}$, the fluctuations of the
order parameter modulus $\rho$ (the so--called longitudinal fluctuations,
which in fact correspond to carrier density fluctuations and
undoubtedly must be taken into account in the very underdoped
region) \cite{foot} are irrelevant and one can safely use the Green's
function (\ref{Green.neutral.Fourier}) of the neutral fermions in the
mean-field approximation (compare with (\ref{Green.standard}))
\begin{equation}
{\cal G}(i \omega_{n}, \mbox{\bf k}) = - \frac{ i \omega_{n}
\hat{I} + \tau_{3} \xi(\mbox{\bf k}) - \tau_{1} \rho}
{\omega_{n}^{2} + \xi^{2}(\mbox{\bf k}) +
\rho^{2}}.
\label{Green.neutral}
\end{equation}
Here $\xi(\mbox{\bf k}) = \mbox{\bf k}^{2}/2m - \mu$ with
$\mbox{\bf k}$  being a 2D vector and
$\rho\equiv\langle\rho(x)\rangle$. Note that in
\cite{Gusynin.JETPLett,Gusynin.JETP}, $\rho(x)$ was treated only
in the mean-field approximation, which means that
fluctuations in both $\rho(x)$ and $\theta(x)$ were neglected,
and therefore a second-order phase transition was obtained at
$T_{\rho}$. However, as stressed in Introduction, experimentally
the formation of the pseudogap phase does not display any sharp
transition and the temperature $T^{\ast}$ observed in various
experiments is to be considered as a characteristic energy scale,
rather than as a temperature where the pseudogap is reduced to zero
\cite{Renner}. We believe that taking the $\rho(x)$ fluctuations
into account may resolve the discrepancy between the experimental
behavior of $T^{\ast}$ and the temperature $T_{\rho}$ introduced
in the theory.

\section{The correlation function for the phase fluctuations}
\setcounter{equation}{0}
\label{sec:phase.green}

As stated above, we expect the phase fluctuations to be
responsible for the difference between properties of the charged and
neutral fermions defined above. The latter are described by the Green's
function (\ref{Green.neutral}), which coincides with the BCS Green's
function (\ref{Green.standard}) only under the assumption that the
phase $\theta$ of the order parameter $\Phi=\rho \exp(i\theta)$ is a
constant and can be chosen to vanish. This is not the case for the 2D
model, where there is a decay of the phase correlations and the
Green's functions of charged and neutral fermions are nontrivially
related via Eq. (\ref{Green.projectors.momentum}). To establish
their relationship, one must know the correlator for the phase
fluctuations. Its calculation is quite straightforward for $T <
T_{\rm BKT}$, while for $T > T_{\rm BKT}$ one can apply the results
of the BKT transition theory \cite{Plischke}.

\subsection{The correlator for $T < T_{\rm BKT}$}
\label{subsec:phase.green}

In the superconducting phase, the free vortex excitations are absent
and the exponential correlator is easily expressed in terms of the
Green's function
\begin{equation}
D_{\theta} (x) =
\langle \theta(x)\theta(0) \rangle
\end{equation}
( here, as above, $x \equiv \tau,\mbox{\bf r}$) via the Gaussian
functional integral
\begin{eqnarray}
&& D_{\alpha \beta}(x)  =\int {\cal D}\theta(x)
\exp \left\{ -\int_{0}^{1/T} \! \! d \tau_{1}
\int  \! \! d^{2}  r_{1}\left[ {1\over2} \theta(x_1)
D_{\theta}^{-1} (x_1)\theta(x_1) + I(x_1)
\theta(x_1) \right] \right\}
\nonumber                  \\
&& = \exp \left[-{1\over 2}
\int_{0}^{1/T} \! \! d \tau_{1} \int_{0}^{1/T} \! \! d \tau_{2}
\int  \! \! d^{2} r_{1} \int  \! \! d^{2} r_{2}
I(\tau_1,{\bf r_1}) D_\theta (\tau_{1} - \tau_{2}, \mbox{\bf r}_{1}
-\mbox{\bf r}_{2}) I(\tau_2,{\bf r_2})\right],
\label{Gaussian}
\end{eqnarray}
with the source
\begin{equation}
I(x_1) = -i \frac{\alpha}{2}
\delta(\tau_{1}- \tau) \delta(\mbox{\bf r}_{1} - \mbox{\bf r}) +
i \frac{\beta}{2} \delta(\tau_{1}) \delta(\mbox{\bf r}_{1}),
\quad (\alpha,\beta=\pm).
\end{equation}
The Green's function
\begin{equation}
D_{\theta}^{-1} (x) =- J(\mu, T, \rho) \nabla^{2}_{r} - K(\mu, T,
\rho) (\partial_{\tau})^{2}
\label{Green.our}
\end{equation}
for this model was found in \cite{Gusynin.JETP}. Note that
the superfluid stiffness $J$ and compressibility $K$ are here the
functions of $\mu$, $T$ and $\rho$, and also that the Green's
function (\ref{Green.our}) includes only the lowest derivatives of
the phase $\theta$. The higher terms are also present in the
expansion, but we neglect them. In the simplest case where
$J(\mu,T,\rho)\sim n_f$, the density of carriers, and
$K(\mu,T,\rho)\sim \mbox{const}$ \cite{Gusynin.JETP}.

Substituting (\ref{Green.our}) into (\ref{Gaussian}), we obtain
\begin{equation}
D_{\alpha \beta}(x)= \exp
\left[ - \frac{T}{4} \sum_{n=-\infty}^\infty
\int \frac{d^2 q}{(2 \pi)^2}
\frac{1- \alpha \beta\cos(\mbox{\bf q} \mbox{\bf r} - \Omega_n \tau)}
{J q^2 + K \Omega^2_n} \right].
\label{correl.1}
\end{equation}
It is easy to see that for zero frequency $\Omega_n = 0$, the
integral in Eq. (\ref{correl.1}) is divergent at ${{\bf q}} = 0$
unless $\alpha =\beta$, and therefore only two terms survive
in the sum over $\alpha, \beta$  in Eq. (\ref{Green.projectors.momentum}),
namely
\begin{equation}
P_{-} {\cal G}(i \omega_{n}, \mbox{\bf k})P_{-} +
P_{+} {\cal G}(i \omega_{n}, \mbox{\bf k})P_{+} =
- \frac{i \omega_{n} \hat I + \tau_{3} \xi(\mbox{\bf k})}
{\omega_{n}^{2} + \xi^{2}(\mbox{\bf k}) + \rho^{2}}.
\label{survive}
\end{equation}
It is important that the terms like $P_{\pm} G(i \omega_{n},
\mbox{\bf k})P_{\mp}$, which are proportional to $\tau_1$ and thus
violate the gauge symmetry, do not contribute to Eq.
(\ref{Green.projectors.momentum}) due to vanishing of the corresponding
$D_{+ -}$ and $D_{- +}$ correlators standing after them. This
explicitly demonstrates that the non-diagonal part of the 2D Green's
function is vanishes at all finite temperatures. Thus, making
use of the Gor'kov equations for the calculation of its diagonal part
and the gap function is questionable. For nonzero correlators, we have
\begin{equation}
D(x) \equiv D_{+ +}(x) = D_{- -}(x) = \exp \left[
- \frac{T}{4} \sum_{n=-\infty}^\infty \int
\frac{q d q d \varphi}{(2\pi)^2}
\frac{1- \cos (q  r \cos \varphi) \cos \Omega_n \tau }
{J q^2 + K \Omega^2_n} \right].
\label{correl.2}
\end{equation}

In what follows, we consider in detail only the static case $\tau=0$.
The restriction to this case is one of the few main
assumptions we use throughout the paper.

The summation over $n$ and the integration over $\varphi$ in
(\ref{correl.2}) can be readily done yielding the following
exponent of (\ref{correl.2})
\begin{equation}
- \frac{1}{16 \pi \sqrt{JK} } \int \limits_0^\infty d q
e^{-q / \Lambda} [1 - J_0(q r)] \tanh {qr_0\over4} ,
\label{exponent}
\end{equation}
where we introduced the scale
\begin{equation}
r_0 = {2\over T} \sqrt{\frac{J}{K}},
\label{a}
\end{equation}
which is a function of the variables used (in the simplest case,
$r_0\sim\sqrt n_f /T$). In (\ref{exponent}), we introduced the
cutoff $\Lambda$ by means of the exponential function. This cutoff
represents the maximal possible momentum in the theory, the
Brillouin momentum.

One can derive from (\ref{exponent}) (see Appendix~\ref{sec:A})
the following asymptotic expressions
\begin{equation}
D(0, \mbox{\bf r}) \sim
\left\{\begin{array}{cc}
\left( \frac{\ds r}{\ds r_0} \right)^{-\frac{\ds T}{\ds 8 \pi J}},
&  r \gg r_0 \gg \Lambda^{-1} \\
\left(\frac{\ds \Lambda r} {\ds 2} \right)^{-\frac{\ds T}{\ds 8 \pi J}},
&  r \gg \Lambda^{-1} \gg r_0.
\end{array} \right.
\label{asymptotic}
\end{equation}
This long-distance behavior governs the physics of
$\theta$-fluctuations that we intend to study in what follows.

We now discuss the meaning of the value $r_0$. Using again the
phase stiffness $J(T=0)$ and  compressibility $K$  from
\cite{Gusynin.JETP} we readily obtain that $r_0 = 2
\sqrt{\epsilon_{F}/m}/T$, which is the single-particle thermal de
Broglie wavelength ($\epsilon_F=\pi n_f/m$ is the Fermi energy). Then,
assuming that $T\sim T_{\rm BKT}$ and taking
$T_{\rm BKT}\simeq\epsilon_F/8$
\cite{Gusynin.JETPLett,Gusynin.JETP}, we can estimate
\begin{equation}
r_0 \sim \frac{16}{\sqrt{\epsilon_{F} m}} = \frac{16\sqrt{2}}{k_F},
\label{a.estimate.Fermi}
\end{equation}
where $k_F$ is the Fermi momentum. The value of $k_F$ for cuprates
is less than the Brillouin momentum $\Lambda$, which is why the
first case in (\ref{asymptotic}) seems to be more relevant.

There is another way to estimate $r_0$: we can use the value $2
\Delta/ T_{c}$, and hence,
\begin{equation}
r_0 \sim \sqrt{2}\pi\frac{2\Delta}{T_c} \xi_{0},
\label{a.estimate.coherence}
\end{equation}
where $\xi_{0} = v_{F}/(\pi \Delta)$ is the BCS coherence length.
This shows that $r_0$ has the meaning of a coherence length, which
appears to be rather natural since the minimal size the phase coherence
region should be of the order of $\xi_{0}$. Since the coherence length
in cuprates is larger than the lattice spacing $\Lambda^{-1}$, we again
obtain that the first case in (\ref{asymptotic}) applies.
Therefore, for $T < T_{\rm BKT}$ and for static fluctuations,
we have that
\begin{equation}
D(\mbox{\bf r}) =
\left( \frac{r}{r_{0}} \right)^{-\frac{\ds T}{\ds 8 \pi J}},
\label{correlator.<}
\end{equation}
where $r_{0} = 16/ \sqrt{\epsilon_{F} m}$.

\subsection{The correlator for $T > T_{\rm BKT}$}

For $T > T_{\rm BKT}$, the expression for static correlator
(\ref{correlator.<}) can be generalized using the well-known
results of the BKT transition theory \cite{Plischke,Pierson},
\begin{equation}
D(\mbox{\bf r}) =
\left( \frac{r}{r_{0}} \right)^{-\frac{\ds T}{\ds 8 \pi J}}
\exp \left( - \frac{r}{\xi_{+}(T)} \right),
\label{correlator.>}
\end{equation}
where
\begin{equation}
\xi_{+}(T) = C \exp
\sqrt{\frac{T_{\rho} - T}{T - T_{\rm BKT}}}
\label{BKT.length}
\end{equation}
is the BKT coherence length and $C$ is a constant whose value
is discussed later. One can consider
Eq.(\ref{correlator.>}) as a general representation for
$D(\mbox{\bf r})$ for both $T > T_{\rm BKT}$ and $T < T_{\rm BKT}$
if the coherence length $\xi_{+}(T)$ is considered to be infinite
for $T < T_{\rm BKT}$. The pre--factor in
Eq.~(\ref{correlator.>}) is related to the longitudinal (spin-wave)
phase fluctuations, while the exponent is responsible for the
transverse (vortex) excitations, which are present only above
$T_{\rm BKT}$. The pre--factor appears to be important
for a non-Fermi liquid behavior discussed in what follows.
Note, however, that the longitudinal phase fluctuations
can be suppressed by the Coulomb interaction \cite{Franz} that
is not included in the present simple model. One further comment is
that while the approximation used to study the vortex fluctuations
in \cite{Franz} is good for $T$ well above $T_{\rm BKT}$, the
form of the correlator $D$ is appropriate for $T$ close to
$T_{\rm BKT}$.

The constant $C$ can be estimated from the condition that
$\xi_{+}(T)$ cannot be much less than the parameter $r_0$ which
is a natural cutoff in the theory and we thus take
$C = r_0/4$ in our numerical calculations $C = r_0/4$.
In any case, for $T\, \rtwid\, T_{\rm BKT}$, where (\ref{BKT.length})
is valid, the value $\xi_{+}(T)$ is large and not very sensitive to
the initial value of $C$.

There also exists a dynamical generalization of (\ref{correlator.>})
proposed from phenomenological backgrounds in \cite{Capezzali},
\begin{equation}
D(t, \mbox{\bf r}) = \exp (- \gamma t)
\left( \frac{r}{r_{0}} \right)^{-\frac{\ds T}{\ds 8 \pi J}}
\exp \left( - \frac{r}{\xi_{+}(T)} \right) .
\label{correlator.dynamic}
\end{equation}
Note that $t$ is the real time and $\gamma$ is the decay constant,
and theherefore (\ref{correlator.dynamic}) is the retarded Green's
function. We hope to consider the more general case of dynamical phase
fluctuations (\ref{correlator.dynamic}) elsewehere.

\subsection{The Fourier transform of $D(\mbox{\bf r})$}

For the Fourier transform (\ref{Green.phase.Fourier}) of
(\ref{correlator.>}), we have
\begin{eqnarray}
D(i \Omega_{n}, \mbox{\bf q}) & = &
\int_{0}^{1 /T} d \tau \int d^2 r\exp\left(i \Omega_{n} \tau
- i \mbox{\bf q} \mbox{\bf r}\right)( r / r_{0} )^{-T/ 8 \pi J}
\exp (- r / \xi_{+}(T) )
\nonumber                         \\
& = & 2\pi \frac{\delta_{n,0}}{T}
r_{0}^{T / 8 \pi J}\int_0^\infty d r r^{1-
T/ 8\pi J}J_0(qr)\exp(- r / \xi_{+}(T)).
\label{Fourier.D1}
\end{eqnarray}
The integral in (\ref{Fourier.D1}) can be calculated
(see, for example, \cite{Gradshteyn}) with the result
\begin{equation}
D(i \Omega_{n}, \mbox{\bf q}) =
\frac{\delta_{n, 0}}{T}
\frac{2\pi r_0^{2(1 - \alpha)} \Gamma(2\alpha)}
{[q^2 + (1/ \xi_{+})^{2}]^{\alpha}} { }_2F_{1} \left(\alpha,
-\alpha+{1\over2}; 1;
\frac{q^2}{q^2 + (1/\xi_{+})^{2}} \right).
\label{Fourier.D2}
\end{equation}
The hypergeometric function $F(a,b;c;z)$ can
be well approximated by a constant since it is slowly varying at
all values of ${\bf q}$. As this constant, we can take the value
of the hypergeometric function at $q=\infty$.  Thus,
\begin{equation}
D(i \Omega_{n}, \mbox{\bf q}) = \frac{\delta_{n, 0}}{T}
A [q^2 + (1/ \xi_{+})^{2} ]^{-\alpha},
\label{Fourier.D.final}
\end{equation}
where
\begin{equation}
A\equiv\frac{4 \pi \Gamma(\alpha) }{ \Gamma(1-\alpha) }
\left( \frac{2}{r_{0}} \right)^{2(\alpha - 1)},
\qquad \alpha\equiv 1 - \frac{T}{16 \pi J}.
\label{A,alpha}
\end{equation}
It should be stressed that for $T > T_{\rm BKT}$, the parameter $\alpha$
quickly deviates from unity as $\epsilon_{F}$ decreases; in other words,
the underdoped region has to reveal highly non-standard properties in
comparison with the overdoped one.

Note that for $\xi_{+}^{-1} = 0$ ($T < T_{\rm BKT}$),
Eq.~(\ref{Fourier.D.final}) is an exact
Fourier transform of the correlator (\ref{correlator.<}).

One should take into account that even for $T < T_{\rm BKT}$, the
propagator (\ref{Fourier.D.final}) does not have the canonical
behavior $\sim 1/q^2$, which is typical, for example, for the
Bogolyubov mode in dimensions $d > 2$. In 2D, the modes with a
propagator $\sim 1/q^2$ would lead to severe infrared singularities
\cite{Coleman}; to avoid them, these modes transform into
softer ones ($\sim 1/q^{2\alpha}$, $\alpha < 1$).

Finally, substituting (\ref{survive}) and (\ref{Fourier.D.final})
in (\ref{Green.projectors.momentum}), we obtain
\begin{equation}
G(i \omega_{n}, \mbox{\bf k}) = - A \int \frac{d^{2} q}{(2 \pi)^2}
\frac{i \omega_{n} + \tau_{3} \xi(\mbox{\bf q})}
{\omega_{n}^2 + \xi^2(\mbox{\bf q}) + \rho^{2}}
\frac{1} {[(\mbox{\bf k} - \mbox{\bf q})^{2} + (1 /
\xi_{+})^{2}]^{\alpha}}.
\label{Green.for.derivation}
\end{equation}
The coincidence of the Matsubara frequency in the left-
and right-hand sides of Eq. (\ref{Green.for.derivation})
is evidently related to the static approximation
used in this paper.
As we see in the next sections, the Green's function
(\ref{Green.for.derivation}), spectral density, and the
density of states can be evaluated exactly.

\section{The derivation of the fermion Green's function}
\setcounter{equation}{0}
\label{sec:green.fermion}

The fermion Green's function can be calculated analytically
if we split the fermion part of (\ref{Green.for.derivation}) as
\begin{equation}
\frac{i\omega_n{\hat I}+\tau_3\xi(\mbox{\bf k})}
{\omega_n^2+\xi^2(\bf k)+\rho^2}=
\frac{A_1}{\xi(\mbox{\bf k})+i\sqrt{\omega_n^2+\rho^2}} +
\frac{A_2}{\xi(\mbox{\bf k})-i\sqrt{\omega_n^2+\rho^2}},
\label{relativistic.spliting}
\end{equation}
where
\begin{equation}
A_{1} = \frac{1}{2} \left(\tau_{3} -
\frac{\omega_{n}}{\sqrt{\omega_{n}^{2} + \rho^{2}}}\right),
\qquad
A_{2} = \frac{1}{2} \left(\tau_{3} +
\frac{\omega_{n}}{\sqrt{\omega_{n}^{2} + \rho^{2}}}\right).
\label{A1,2.Matsubara}
\end{equation}
Using the representations
\begin{equation}
{1\over a\pm i b} = \mp i\int\limits_0^\infty ds
\exp{[\pm i s(a \pm i b)]}\,,
\label{representation.Im}
\end{equation}
\begin{equation}
\frac{1}{c^\alpha} =
{1\over\Gamma(\alpha)}\int\limits_0^\infty dtt^{\alpha-1}e^{-ct}
\label{representation.Re}
\end{equation}
and taking (\ref{relativistic.spliting}) into account, we can
rewrite (\ref{Green.for.derivation}) as
\begin{eqnarray}
&& G(i \omega_{n}, \mbox{\bf k}) = \frac{i A}{\Gamma(\alpha)}
\int_{0}^{\infty} ds \int_{0}^{\infty} dt t^{\alpha - 1}
e^{-\xi_{+}^{-2} t - s \sqrt{\omega_{n}^{2} + \rho^{2}}} \times
\nonumber\\
&& \int \frac{d^{2} q}{(2 \pi)^{2}} \left\{ A_{1} \exp \left[ i s
\frac{q^{2}}{2m} - i\mu s - (\mbox{\bf k}
-\mbox{\bf q})^{2} t \right] -
A_{2} \exp \left[- i s \frac{q^{2}}{2m} + i\mu s - (\mbox{\bf k}
-\mbox{\bf q})^{2} t \right] \right\}.
\label{Green.integral}
\end{eqnarray}
Note that the special form of the integral representation
(\ref{representation.Im}) (compare with representation
(\ref{representation.Re})) guarantees that the Gaussian integral
over $q$ is well-defined independently of the sign of
$\xi(\mbox{\bf q})={\bf q}^2/2m -\mu$. Now the Gaussian integration
over momenta $q$ in (\ref{Green.integral}) can be done explicitly:
\begin{eqnarray}
&& G(i \omega_{n}, \mbox{\bf k}) = \frac{i A}{4 \pi \Gamma(\alpha)}
\int_{0}^{\infty} ds \int_{0}^{\infty} dt t^{\alpha - 1}
e^{- \xi_{+}^{-2} t - s \sqrt{\omega_{n}^{2} + \rho^{2}}} \times
\nonumber\\
&& \left[ \frac{A_{1}}{t - is / 2m}
\exp \left( i\frac{{\bf\ k}^2}{2m}\frac{ s t}{t - is / 2m} -
i \mu s \right)- \frac{A_{2}}{t + is / 2m}
\exp \left( - i\frac{{\bf k}^2}{2m}\frac{ s t}{t + is / 2m}
+ i \mu s \right)\right].
\end{eqnarray}

Changing the variables as $s \to 2m s$ and further as $t \to st$,
we can integrate over $s$ with the result
\begin{eqnarray}
G(i \omega_{n}, \mbox{\bf k}) = && \frac{i m A}{2 \pi}
\int_{0}^{\infty} dt t^{\alpha - 1}
\left\{ \frac{A_{1} (t - i)^{\alpha - 1}}
{\left[ \xi_{+}^{-2} t (t - i) + 2m \sqrt{\omega_{n}^{2} + \rho^{2}} (t-i)
- i t{\bf k}^{2} + 2 i m \mu (t - i)
\right]^{\alpha}} \right.
\nonumber\\
&& - \left.
\frac{A_{2} (t + i)^{\alpha - 1}}
{\left[ \xi_{+}^{-2} t (t + i) + 2m \sqrt{\omega_{n}^{2} +
\rho^{2}} (t+i) + i t {\bf k}^{2} - 2 i m \mu (t + i)\right]^{\alpha}}
\right\}.
\label{Green.replaced}
\end{eqnarray}

In the general case where
$\xi_{+}^{-1} \neq 0$, the denominator of (\ref{Green.replaced})
is quadratic in $t$ and some further transformations are needed.
Replacing $t \to - i u$ and expanding the quadratic polynomial in
the denominator, we have
\begin{equation}
G(i \omega_{n}, \mbox{\bf k})
= -\frac{A m \xi_{+}^{2\alpha}}{2\pi} \left\{
\int_{0}^{i \infty} du \frac{A_1 u^{\alpha - 1} (u + 1)^{\alpha - 1}}
{[(u + u_1) (u + u_2)]^{\alpha}} +
\int_{0}^{-i \infty} du \frac{A_2 u^{\alpha - 1} (u + 1)^{\alpha - 1}}
{[(u + \tilde u_1) (u + \tilde u_2)]^{\alpha}} \right\},
\label{Green.imaginary.limits}
\end{equation}
where
\begin{eqnarray}
u_1 & = & m \xi_{+}^{2} \left(
\frac{k^2\xi_{+}^2+1}{2m\xi_{+}^2} - \mu + i \sqrt{\omega_{n}^{2} +
\rho^{2}} +\sqrt D \right),
\nonumber             \\
u_2 & = & m \xi_{+}^{2} \left(
\frac{k^2\xi_{+}^2+1}{2m\xi_{+}^2} - \mu + i \sqrt{\omega_{n}^{2} +
\rho^{2}} -\sqrt D \right)
\label{roots.Matsubara}
\end{eqnarray}
with
\begin{equation}
D\equiv \left( \frac{k^2\xi_{+}^2+1}{2m\xi_{+}^2} - \mu + i
\sqrt{\omega_{n}^{2} +\rho^{2}} \right)^2 + \frac{2}{m \xi_{+}^{2}}
(\mu - i \sqrt{\omega_{n}^{2} + \rho^{2}})
\label{D.Matsubara}
\end{equation}
and
\begin{equation}
\tilde u_i = u_i(\sqrt{\omega_{n}^{2} + \rho^{2}} \to -
\sqrt{\omega_{n}^{2} + \rho^{2}}).
\end{equation}

We can verify from (\ref{roots.Matsubara}) that $\mbox{Re} u_i > 0$
for $\mu < 0$, and therefore, we can rotate the integration contour
to the real axis:
\begin{equation}
G(i \omega_{n}, \mbox{\bf k}) =
-\frac{A m \xi_{+}^{2\alpha}}{2\pi}
\left\{
\int_{0}^{\infty} du \frac{A_1 u^{\alpha - 1} (u + 1)^{\alpha - 1}}
{[(u + u_1) (u + u_2)]^{\alpha}} +
(\sqrt{\omega_{n}^{2} +
\rho^{2}} \to -\sqrt{\omega_{n}^{2} + \rho^{2}})
\right\}.
\label{Green.real.limits}
\end{equation}
The integral representation (\ref{Green.real.limits}) can then be
analytically continued to $\mu > 0$. The change of the variable $z
= u/(u+1)$ allows Eq.~(\ref{Green.real.limits}) to be expressed in terms
of Appell's function \cite{Bateman}
\begin{equation}
F_{1} (\alpha, \beta, \beta^{\prime}, \gamma; x, y) =
\frac{\Gamma(\gamma)}{\Gamma(\alpha) \Gamma(\gamma - \alpha)}
\int_{0}^{1} \frac{z^{\alpha-1} (1 - z)^{\gamma-\alpha-1}}
{(1 - zx)^{\beta} (1 - zy)^{\beta^{\prime}}} dz,
\label{Appell.definition}
\end{equation}
and hence,
\begin{eqnarray}
G(i \omega_{n},\mbox {\bf
k})=-\frac{Am\xi_{+}^{2\alpha}}{2\pi\alpha} &&\left[\frac{A_1
}{(u_1 u_2)^{\alpha}} F_1
\left(\alpha, \alpha, \alpha; \alpha + 1;
\frac{u_1 - 1}{u_1}, \frac{u_2 - 1}{u_2} \right) \right.
\nonumber\\
&&  + \left. (\sqrt{\omega_{n}^{2} +
\rho^{2}} \to -\sqrt{\omega_{n}^{2} + \rho^{2}}) \right].
\label{Green.final.Matsubara}
\end{eqnarray}

For $T < T_{\rm BKT}$, the BKT coherence length is infinite
($\xi_{+}^{-1} = 0$), which means $(u_1-1)/u_1 = 1$ in
the first argument of the Appell's function. This allows us to
apply the reduction formula \cite{Bateman}
\begin{equation}
F_1(\alpha, \beta, \beta^{\prime}, \gamma;x,1) =
\frac{\Gamma(\gamma) \Gamma(\gamma - \alpha - \beta^{\prime})}
{\Gamma(\gamma - \alpha) \Gamma(\gamma - \beta^{\prime})}
\, \, {}_2F_{1} (\alpha, \beta; \gamma-\beta^{\prime};x)
                       \label{AppelltoHyper.in1}
\end{equation}
and express the result via the hypergeometric function
\begin{eqnarray}
G(i \omega_{n}, \mbox{\bf k})&=& -
\Gamma^{2}(\alpha) \left(\frac{2}{m r_{0}^{2}} \right)^{\alpha - 1}
\left\{\frac{A_1}{[- (\mu - i \sqrt{\omega_{n}^{2} + \rho^{2}})]^\alpha}
{ }_2F_{1} \left(\alpha, \alpha; 1;
\frac{k^2/2m}{\mu - i \sqrt{\omega_{n}^{2} + \rho^{2}})} \right)
\right.
\nonumber\\
&& \left. +
\frac{A_2}{[- (\mu + i \sqrt{\omega_{n}^{2} + \rho^{2}})]^\alpha}
{ }_2F_{1} \left(\alpha, \alpha; 1;
\frac{k^2/2m}{\mu + i \sqrt{\omega_{n}^{2} + \rho^{2}})} \right)
\right\},
\label{Green.<.final}
\end{eqnarray}
where we inserted the value of $A$ from (\ref{A,alpha}).

This completes our derivation of the temperature fermion Green's
function.

\subsection{The retarded fermion Green's function}

To obtain the spectral density, we need to
obtain the retarded real-time Green's function from the temperature
Green function by means of analytical continuation $i \omega_n \to
\omega + i 0$, and where $\sqrt{\omega_{n}^{2} + \rho^{2}} \to i
\sqrt{\omega^{2} -
\rho^{2}}$. This results in the following rules (compare with
(\ref{A1,2.Matsubara}), (\ref{roots.Matsubara}),
(\ref{D.Matsubara}))
\begin{equation}
A_{1} \to {\cal A}_{1} = \frac{1}{2} \left(\tau_{3} +
\frac{\omega}{\sqrt{\omega^{2} - \rho^{2}}}\right),
\qquad
A_{2} \to {\cal A}_{2} = \frac{1}{2} \left(\tau_{3} -
\frac{\omega}{\sqrt{\omega^{2}-\rho^{2}}}\right)\,,
\label{A1,2.real}
\end{equation}
\begin{eqnarray}
u_1 \to v_1 & = & m \xi_{+}^{2} \left(
\frac{k^2\xi_{+}^2+1}{2m\xi_{+}^2} - \mu - \sqrt{\omega^{2} -
\rho^{2}} +\sqrt {\cal D} \right),
\nonumber             \\
u_2 \to v_2 & = & m \xi_{+}^{2} \left(
\frac{k^2\xi_{+}^2+1}{2m\xi_{+}^2} - \mu - \sqrt{\omega^{2} -
\rho^{2}} -\sqrt {\cal D} \right),
\label{roots.real}
\end{eqnarray}
with
\begin{equation}
D \to {\cal D} =
\left( \frac{k^2\xi_{+}^2+1}{2m\xi_{+}^2} - \mu - \sqrt{\omega^{2} -
\rho^{2}} \right)^2 + \frac{2}{m \xi_{+}^{2}}
(\mu + \sqrt{\omega^{2} - \rho^{2}})
\label{D.real}
\end{equation}
and
\begin{equation}
\tilde v_i = v_i(\sqrt{\omega^{2} - \rho^{2}} \to -
\sqrt{\omega^{2} - \rho^{2}}).
\end{equation}
For the retarded Green's function we thus have
\begin{eqnarray}
G(\omega, \mbox{\bf k}) = -\frac{A m \xi_{+}^{2\alpha}}{2\pi\alpha}
&&
\left\{ \frac{{\cal A}_1 }{(v_1 v_2)^{\alpha}} F_1
\left(\alpha, \alpha, \alpha; \alpha + 1;
\frac{v_1 - 1}{v_1}, \frac{v_2 - 1}{v_2} \right) \right.
\nonumber          \\
&&  + \left. (\sqrt{\omega^{2} - \rho^{2}} \to
-\sqrt{\omega^{2} - \rho^{2}}) \right\}.
\label{Green.final.real}
\end{eqnarray}
It is easy to see that
\begin{equation}
v_1 v_2 = - 2m \xi_{+}^{2} (\mu + \sqrt{\omega^{2} - \rho^{2}}).
\label{product}
\end{equation}

We now discuss the condition under which the imaginary part of
$G(\omega + i 0, \mbox{\bf k})$ is nonvanishing.

For $|\omega| < \rho$, we can see that $\tilde v_1 = v_1^\ast$,
$\tilde v_2 = v_2^\ast$, and therefore $G(\omega, \mbox{\bf k})$ is real
and $\mbox{Im} G(\omega+ i 0, \mbox{\bf k}) =0$. The case where
$|\omega|> \rho$ is more complicated. It follows from the
Appell's function transformation property \cite{Bateman}
\begin{equation}
F_1 (\alpha, \beta, \beta^{\prime}, \gamma; x, y) =
(1 - x)^{-\alpha}
F_1 \left(\alpha, \gamma-\beta-\beta^{\prime}, \beta^{\prime},
\gamma; \frac{x}{x - 1}, \frac{y - x}{1 - x} \right).
\label{Applel.property}
\end{equation}
that for real $x$ and $y$, the function $F_1$ becomes complex if $x
> 1$ or/and $y > 1$. This implies that $G(\omega, \mbox{\bf k})$
has an imaginary part if $v_1 < 0$ or/and $v_2 < 0$. Looking at the
expressions (\ref{roots.real}) for $v_1$ and $v_2$, we can see that
$v_1$ is always positive, while $v_2$ may be negative. This means
that $G(\omega, \mbox{\bf k})$ has a nonvanishing imaginary part
if $v_1 v_2 < 0$.  Using (\ref{product}), the condition for the
existence of a nonzero imaginary part of $G(\omega, \mbox{\bf k})$
can then be written as $\mu + \sqrt{\omega^{2} - \rho^{2}} > 0$.

\subsection{The branch cut structure of $G(\omega, \mbox{\bf k})$ and
a non-Fermi liquid behavior}

We now consider the retarded fermion Green's function
(\ref{Green.<.final}) for $T < T_{\rm BKT}$. Applying the
analytic continuation rules from the previous subsection to
Eq. (\ref{Green.<.final}), we obtain
\begin{eqnarray}
G(\omega, \mbox{\bf k}) & = &
-  \Gamma^{2}(\alpha) \left(\frac{2}{m r_{0}^{2}} \right)^{\alpha - 1}
\left[ \frac{{\cal A}_1}
{[- (\mu + \sqrt{\omega^{2} - \rho^{2}})]^\alpha} { }_2F_{1}
\left(\alpha, \alpha; 1;
\frac{k^2/2m}{\mu + \sqrt{\omega^{2} - \rho^{2}})} \right)\right.
\nonumber\\
&& \left. +\frac{{\cal A}_2} {[- (\mu - \sqrt{\omega^{2} -
\rho^{2}})]^\alpha} { }_2F_{1}
\left(\alpha, \alpha; 1;
\frac{k^2/2m}{\mu - \sqrt{\omega^{2} - \rho^{2}})} \right)\right].
\label{Green.<.final.real}
\end{eqnarray}

Near the quasiparticle peaks where $\omega \approx \pm E(\mbox{\bf
k})$, the arguments of the hypergeometric function in
(\ref{Green.<.final.real}) are close to 1. One can consider, for
instance, the first hypergeometric function, then
\begin{equation}
z_{1} \equiv \frac{k^2/2m}{\mu + \sqrt{\omega^{2} - \rho^{2}}}
\simeq 1.
\end{equation}
Using the relation between the hypergeometric functions
\cite{Bateman}
\begin{eqnarray}
{}_2F_1(a,b;c;z) & = &
\frac{\Gamma(c) \Gamma(c-a-b)}{\Gamma(c-a)\Gamma(c-b)}
{}_2F_1(a, b; a+b+1-c; 1-z)
\nonumber\\
& + & \frac{\Gamma(c)\Gamma(a+b-c)}{\Gamma(a)\Gamma(b)}(1-z)^{c-a-b}
{}_2F_1(c-a,c-b;c+1-a-b;1-z)\,,
\end{eqnarray}
we obtain that near $z_1 \simeq 1$,
\begin{eqnarray}
G(\omega, \mbox{\bf k})  && \sim-\Gamma^{2}(\alpha)\hspace{-1mm}
\left(\frac{2}{m r_{0}^{2}}\right)^{\alpha - 1}\hspace{-7mm}
\frac{{\cal A}_1}{[- (\mu + \sqrt{\omega^{2} - \rho^{2}})]^\alpha}
\hspace{-1mm}
\left\{  \frac{\Gamma(1-2\alpha)}{\Gamma^{2}(1-\alpha)}\hspace{-0.5mm}
+\hspace{-0.5mm}
\frac{\Gamma(2\alpha-1)}{\Gamma^{2}(\alpha)}\frac{1}{(1-z_1)^{
2\alpha - 1}}\right\}.
\label{cut.<}
\end{eqnarray}
It can be seen that the expression for the Green's function obtained is
evidently a nonstandard one: besides containing a branch cut, it
clearly displays its non-pole character. The latter in its turn
corresponds to a non-Fermi liquid behavior of the system as a
whole. It must be stressed that non-Fermi liquid peculiarities
are tightly related to the charge (i.e., observable) fermions only,
because the Green's function (\ref{Green.neutral}) of neutral
fermions has a typical (pole type) BCS form. In addition, it follows
from (\ref{cut.<}) that new properties appear as a consequence of the
$\theta$-particle presence (leading to $\alpha\neq1$), and because the
parameter $\alpha$ is a function of $T$ (see (\ref{A,alpha})), the
non-Fermi liquid behavior is developed with temperature increase
and is preserved until $\rho$ vanishes.

It is interesting that in Anderson's theory \cite{Anderson}, it was
postulated that the Fermi liquid theory is broken down in the normal
state as a result of strong correlations. Here, we started
from the Fermi liquid theory and found that it is broken down due
to strong phase fluctuations. As suggested in \cite{Anderson},
the non-Fermi liquid behavior may lead to the suppression of
the coherent tunneling between layers, which in turn confines
carriers in the layers and leads to the strong phase fluctuations.
In contrast to \cite{Anderson}, however, our model predicts the
restoration of the Fermi liquid behavior as $T$ decreases, since
$\alpha \to 1$ as $T \to 0$ (see the discussion in
Sec.~\ref{subsec:results} item~4.).

The $T=0$ limit can also be obtained as follows.
Strictly speaking, one cannot estimate the value of
$r_{0}$ in the limit as $T \to 0$ in (\ref{Green.<.final.real}) via
Eq. (\ref{a.estimate.Fermi}), because the substitution of
$T_{\rm BKT}\simeq \epsilon_F/8$ in (\ref{a}) is
not valid in this case. However, this is not essential because
$T/ 8\pi J \to 0$, so that the correlator (\ref{correlator.<}),
$D(\mbox{\bf r}) \to 1$, which evidently means the formation of
a long-range order in the system.  Furthermore, the value of
$\alpha$ in (\ref{A,alpha}) goes to 1 as $T \to 0$, and the
hypergeometric function in (\ref{Green.<.final.real})
reduces to the geometrical series,
\begin{equation}
{ }_2F_{1} (1, 1; 1; z) = \frac{1}{1-z}.
\label{geometrical}
\end{equation}
Therefore, inserting (\ref{geometrical}) in (\ref{Green.<.final.real}),
we obtain the standard BCS expression
\begin{equation}
G_{1 1}(\omega, \mbox{\bf k}) =
\frac{\omega + \xi(\mbox{\bf k})}
{\omega^{2} - \xi^{2}(\mbox{\bf k}) - \rho^{2}}\,,
\label{BCS.Green11}
\end{equation}
for the diagonal component $G_{11}(\omega, \mbox{\bf k})$ of the
Nambu-Gor'kov Green's function $G(\omega, \mbox{\bf k})$.

Evidently Eq.(\ref{BCS.Green11}) results in the standard BCS spectral
density \cite{Schrieffer} with two $\delta$-function peaks
\begin{equation}
A(\omega, \mbox{\bf k}) =
\frac{1}{2} \left[1 + \frac{\xi(\mbox{\bf k})}{E(\mbox{\bf k})} \right]
\delta(\omega - E(\mbox{\bf k})) +
\frac{1}{2} \left[1 - \frac{\xi(\mbox{\bf k})}{E(\mbox{\bf k})} \right]
\delta(\omega + E(\mbox{\bf k})),
\label{BCS.spectral.density}
\end{equation}
where $E(\mbox{\bf k}) = \sqrt{\xi^{2}(\mbox{\bf k}) + \rho^{2}}$.
To recover the nondiagonal components of $G$, one has to restore the
correlators $D_{- +}(\mbox{\bf r})$ and $D_{+
-}(\mbox{\bf r})$ that were omitted in Sec.~\ref{subsec:phase.green}.

\section{The spectral density of the fermion Green's function}
\setcounter{equation}{0}
\label{sec:spectral}

As is well known,  \cite{Schrieffer}, the spectral features of any
system are entirely controlled by its spectral density
\begin{equation}
A(\omega, \mbox{\bf k}) = -
\frac{1}{\pi} \mbox{Im} G_{11}(\omega +i 0, \mbox{\bf k})\,,
\label{spectral.definition}
\end{equation}
which, for example, for cuprates is measured in ARPES experiments
(see \cite{Campuzano}). This function defines the spectrum
anisotropy, the presence of a gap, the DOS, etc. In what follows, we
calculate $A(\omega,{\bf k})$ for the Green's function obtained
above.

\subsection{Analytical expression for the spectral density}

 For $v_1 > 0$ and $v_2 < 0$, the retarded fermion Green's
function (\ref{Green.final.real}) can be rewritten
(see Appendix~\ref{sec:B}) as
\begin{eqnarray}
G(\omega, \mbox{\bf k}) && =
- \frac{A m \xi_{+}^{2\alpha}}{2 \pi}
\left\{ {\cal A}_1 \left[ \frac{(-1)^{\alpha} \Gamma(\alpha)
\Gamma(1-\alpha)}{ [v_1 (1-v_2)]^{\alpha}}
{}_2F_{1} \left(\alpha, \alpha; 1;
\frac{v_2 (1-v_1)}{v_1(1-v_2)} \right) + \right. \right.
\nonumber                    \\
&& \left. \left.  \frac{1}{|v_2|}
\frac{\Gamma(1-\alpha)}{\Gamma(2-\alpha)}
F_1 \left(1, \alpha, 1-\alpha; 2-\alpha;
\frac{v_1}{v_2}, \frac{1}{u_2} \right) \right]  +
(\sqrt{\omega^{2} - \rho^{2}} \to
-\sqrt{\omega^{2} - \rho^{2}}) \right\}.
\label{Green.real.spectral}
\end{eqnarray}
Then, according to (\ref{spectral.definition}) the spectral density
for the Green's function (\ref{Green.real.spectral}) has the form
\begin{eqnarray}
& &A(\omega, \mbox{\bf k})=
\frac{A m \xi_{+}^{2\alpha}\sin(\pi\alpha)}{2 \pi^{2}}
\mbox{sgn}\omega\,\theta(\omega^2 - \rho^2)
\left[({\cal A}_1)_{11}  \frac{\Gamma(\alpha) \Gamma(1-\alpha)}
{[v_1(1-v_2)]^{\alpha}}\right.\nonumber\\ & &\times\left.
{}_2F_{1}\left(\alpha,\alpha; 1;\frac{v_2(1-v_1)}{v_1(1-v_2)}
\right)\theta(\mu + \sqrt{\omega^{2} - \rho^{2}})
 - (\sqrt{\omega^{2} - \rho^{2}} \to
-\sqrt{\omega^{2} - \rho^{2}}) \right].
\label{spectral.density1}
\end{eqnarray}
Using the quadratic transformation for the hypergeometric function
\cite{Bateman}
\begin{equation}
{}_2F_1(a, b; a-b+1;z) = (1 - z)^{-a} {}_2F_1
\left(\frac{a}{2}, -b+\frac{a+1}{2}; 1+a-b; -\frac{4z}{(1-z)^2} \right)\,,
\end{equation}
the expression (\ref{A,alpha}) for $A$,
Eqs.~(\ref{roots.real}), and (\ref{D.real})
we finally obtain
\begin{eqnarray}
A(\omega, \mbox{\bf k}) =&&
\frac{\Gamma(\alpha)}{\Gamma(1-\alpha)}
\left( \frac{2}{m r_{0}^{2}} \right)^{\alpha-1}
\mbox{sgn} \omega\,\theta(\omega^2 - \rho^2)\times\nonumber\\
&&\left[\frac{({\cal A}_1)_{11}}{{\cal D}^{\alpha/2}}{}_2F_1
\left(\frac{\alpha}{2},\frac{1-\alpha}{2}; 1; -4
\frac{\frac{k^2}{2m}(\mu + \sqrt{\omega^2- \rho^2})}{{\cal D}}\right)
\theta(\mu + \sqrt{\omega^2 - \rho^2})\right.\nonumber\\
&&\left.- (\sqrt{\omega^{2} - \rho^{2}} \to
-\sqrt{\omega^2 - \rho^2}) \right],
\label{spectral.density.final}
\end{eqnarray}
where the chemical potential $\mu$ can be, in principle, determined
from the equation that fixes the carrier density \cite{Gusynin.JETP}.
Here, however, we assume that the carrier density is sufficiently
high and $\mu = \epsilon_{F}$.

In the BCS theory, $A(\omega, \mbox{\bf k})$ given by
Eq.~(\ref{BCS.spectral.density}) consists of two pieces that are the
spectral weights of adding and removing a fermion from the system
respectively. Note that our splitting of $A(\omega, \mbox{\bf k})$
is different since each term in (\ref{spectral.density.final})
corresponds to both the addition and the removal of a fermion.

In the next subsections, we verify the sum rule for
(\ref{spectral.density.final}), plot it for different
temperatures, and discuss the results.

\subsection{The sum rule for the spectral density}

It is well known that for the exact Green's function $G(\omega,
\mbox{\bf k})$, the spectral function (\ref{spectral.definition})
must satisfy the sum rule
\begin{equation}
\int_{- \infty}^{\infty} d \omega A(\omega, \mbox{\bf k}) = 1.
\label{sum.rule}
\end{equation}
The Green's function calculated in (\ref{Green.final.real}) is, of
course, approximate. This is related to the use of
the long-distance asymptotic behavior (\ref{asymptotic}) of the phase
correlator (\ref{correl.2}). This means that its Fourier
transform (\ref{Fourier.D.final}) is, strictly speaking, valid
for small $\mbox{\bf k}$ only, while we have integrated our
expressions to the infinity.  Another approximation that we have made
was the restriction to the static phase fluctuations.  Thus, it
is important to check whether the sum rule (\ref{sum.rule}) is
satisfied with sufficient accuracy.

It is remarkable that for (\ref{spectral.density.final}), the sum
rule (\ref{sum.rule}) can be tested analytically with the help of the
techniques used in calculating $N(\omega)$ in Appendix~\ref{sec:C}.
We obtain
\begin{equation}
\int_{- \infty}^{\infty} d \omega A(\omega, \mbox{\bf k}) = \frac{\Gamma
(\alpha)}{\Gamma(2-\alpha)}.
\label{devsumrule}
\end{equation}

The numerical value of the integral at the temperatures
of interest can be estimated as follows.
Taking the phase stiffness $J = 2/\pi T_{\rm BKT}$
at $T=T_{\rm BKT}$, the value $\alpha$ from
(\ref{A,alpha}) is given by
\begin{equation}
\alpha \simeq 1 - \frac{1}{32} \frac{T}{T_{\rm BKT}}\,,
\qquad T \sim T_{\rm BKT}
\label{alpha.estimation}
\end{equation}
for $T$ close to $T_{\rm BKT}$.
In particular, $\alpha(T=T_{\rm BKT}) =31/32$ gives the following
estimate for the right-hand side of (\ref{devsumrule}),
$\Gamma(\alpha)/\Gamma(2-\alpha) \simeq 1.037$. This shows that
for $T \sim T_{\rm BKT}$, the spectral density
(\ref{spectral.density.final}) is reasonably good at the temperatures of
interest.

The parameter $\alpha$ can however differ strongly from unity
at $T>T_{\rm BKT}$ and in the underdoped regime.

\subsection{Results for the spectral density}
\label{subsec:results}

The plots of the spectral density $A(\omega, \mbox{\bf k})$
given by (\ref{spectral.density.final}) at temperatures
below and above $T_{\rm BKT}$ are presented in Figs.1--3. To draw
these plots, we used the value of $\alpha$ from Eq.
(\ref{alpha.estimation}) and the mean-field value of $\rho$
obtained from the corresponding equation in
\cite{Gusynin.JETPLett,Gusynin.JETP}. From these
figures and our analytical expressions, we can infer the
following results:

\begin{enumerate}

\item  For $T < T_{\rm BKT}$ (the case presented on Fig.~1), there
are two highly pronounced quasiparticle peaks at $\omega = \pm
E(\mbox{\bf k})$. They are simply related to the contribution of
zeros of $\cal D$ (see Eq. (\ref{D.real})) to
$A(\omega, \mbox{\bf k})$.

\item We also observe two peaks at $\omega = \pm \rho$
when $k\neq k_F$ (for $k=k_F$, the value $E(\mbox{\bf k}_{F}) =
\rho$, so that the two sets of peaks coincide).
One can check that the divergence at these points is weaker than at
the former peaks at $\omega = \pm E(\mbox{\bf k})$.  In fact, these
peaks are the result of the static and nointeracting
approximation for the phase fluctuations used here. They are essential
to satisfy the sum rule (\ref{sum.rule}).

If the dynamical fluctuations are taken into account, it is clear
that the ``external'' frequency $\omega$ in $A(\omega, \mbox{\bf k})$
is different from the ``internal'' frequency in ${\cal A}_1,
{\cal A}_2$ (see the discussion after Eq.
(\ref{Green.for.derivation}) and compare it with Eq.
(\ref{Green.projectors.momentum})). We believe that this additional
summation over the``internal'' frequency (which is present if the
dynamical fluctuations are considered) would considerably smear
these peaks, moving the excess of the spectral weight inside the
gap. This assumption is supported by the results of
\cite{Capezzali} (see item~3 below).
The same effect can also be reached when the interaction between
the charge and spin degrees of the freedom is taken into account
\cite{Franz}.
Note also that the full cancellation of these peaks
takes place in the $T =0$ case given by Eq. (\ref{BCS.Green11}).

\item For $\omega < |\rho|$,  we have $A(\omega, \mbox{\bf k}) =0$
and a gap exists at all $T$ (including $T>T_{\rm BKT}$). This result
is also a consequence of the static approximation used above. The
dynamical fluctuations should fill the empty region resulting in
the pseudogap formation in the normal state.
Indeed, a filling of the gap was obtained in a related
calculation \cite{Capezzali}
where the correlator
$\langle\exp(i\theta(\mbox{\bf r}, t)\exp(-i\theta(0))\rangle$
(which differs from (\ref{correlator.dynamic}) only by the
factor $1/2$ muliplying the phase), which includes the dynamical
phase fluctuations, was used in the numerical calculation
of the self-energy of fermions and in the subsequent extraction of
the spectral function from the fermion Green's function.

In the approximation used in the present paper, the spin and charge
degrees of freedom are decoupled (see Eq. (\ref{Green.splited})).
However, this coupling can be included at the next stage of
approximation and also leads to a pseudogap filling. Indeed, using
the special form of the scattering rate proposed in \cite{Norman},
it was obtained in \cite{Franz} that $A(\omega, \mbox{\bf k}) \neq 0$
even for $\omega<|\rho|$. On the other hand, as stated above,
there also are indications \cite{Capezzali} that a filling of the gap
can be obtained by considering the dynamical phase fluctuations only.
At present, it is not clear which of these gap filling mechanisms
plays the main role; this is the subject of our current
investigations.

\item The main peaks at $\omega = \pm E(\mbox{\bf k})$
have a finite temperature dependent width which is, of
course, related to the spin-wave (longitudinal) phase fluctuations.
As $T \to 0$, the width goes to zero, but this limit cannot be
correctly derived from (\ref{spectral.density.final}) because this is
an ordinary function, while the BCS spectral density
(\ref{BCS.spectral.density}) is a distribution. The correct limit
can, however, be obtained for the integral of $A(\omega, \mbox{\bf k})$
(see Sec.~\ref{sec:DOS}, where the density of states is discussed).
This sharpening of the peaks
with decreasing $T$ in the superconducting state was experimentally
observed \cite{Campuzano} and represents a striking difference
from the BCS ``pile-up'' (\ref{BCS.spectral.density}) which are
present for all $T < T_{c}$.

It was pointed out in \cite{Franz} that the broadening
of the spectral function caused by these fluctuations can be greater
than the experimental data permits. This leads \cite{Franz} to the
conclusion that the spin-wave phase fluctuations are probably
suppressed by the Coulomb interaction.

\item For $T > T_{\rm BKT}$ (see Figs.~2,3), one can see that the
quasiparticle peaks at $\omega \approx \pm E(\mbox{\bf k})$ are
less pronounced as the temperature increases. Indeed, the
value of $A(\omega, \mbox{\bf k})$ at $\omega = \pm E(\mbox{\bf
k})$ is, in contrast to the case where $T < T_{\rm BKT}$, already finite.
This is caused by the fact that ${\cal D} \neq 0$ since $\xi_{+}$
is already finite due to the influence of the vortex
fluctuations. As the temperature is increases further, $\xi_{+}$
decreases, so that the quasiparticle peaks disappear (compare
Figs.~2 and 3). This behavior qualitatively reproduces the ARPES
studies of the cuprates for the anti-node direction \cite{Campuzano}
(see also \cite{Norman.unpub}) which show that the quasiparticle
spectral function broadens dramatically when passing from the
superconducting to normal state.

\item It is important to stress that due to a very smooth
dependence of $\xi_{+}^{-1}$ on $T$ (see Eq. (\ref{BKT.length})) as
the temperature varies from $T < T_{\rm BKT}$ to $T > T_{\rm BKT}$,
there is no sharp transition at the point $T = T_{\rm BKT}$.
There is a smooth evolution of the superconducting
(excitation) gap $\Delta_{\rm SC} = \rho$ into the gap $\Delta_{\rm
PG}$, which also is equal to $\rho$ and in fact can be called a
pseudogap because the system is not superconducting at  $T>T_{\rm BKT}$.
This qualitatively fits the experiment
\cite{Campuzano,Renner,Norman.unpub} and appears to be completely
different from the BCS theory \cite{Schrieffer}, where the gap
vanishes at $T=T_{c}$. As was already mentioned, the gap obtained
at $T>T_{\rm BKT}$ takes place in the static approximation only and
begins to be filled after dynamical fluctuations are taken into
account (see, for example, \cite{Capezzali}).

\item Again for $T > T_{\rm BKT}$,  one has
$A(\omega, \mbox{\bf k}) =0$ when $|\omega| < \rho$, and we expect
the gapped region to be filled by the dynamical phase fluctuations
\cite{Capezzali}.
We predict, however, an essential difference between the filling of the
gap at $T > T_{\rm BKT}$ and $T < T_{\rm BKT}$.  Indeed, due to the
presence of the vortices above $T_{\rm BKT}$, the value of the decay
constant $\gamma$ in Eq.  (\ref{correlator.dynamic}) should be much
larger than for $T < T_{\rm BKT}$. This and a nonzero value of
$\xi_{+}^{-1}$ above $T_{\rm BKT}$ may explain the break at $T = T_{c}$
in the scattering rate $\Gamma_{1}$ introduced in \cite{Norman}.
In general, it is interesting to establish a correspondence between
the phenomenological parameters, $\Gamma_{1}$ and $\Gamma_{0}$
introduced in \cite{Norman} and the vortex parameters $\xi_{+}$
and $\gamma$ used here. Note, however, that this correspondence
cannot be simple because of the non-pole character of the Green's
function derived here.

As mentioned above (see item~3), the filling of the gap
due to dynamical phase fluctuations is not the only possible
mechanism for filling and the presence of vortices
above $T_{\rm BKT}$ can be taken into account via coupling
the spin and charge degrees of freedom \cite{Franz}. It could also
be that both these mechanisms are physically equivalent, since they
relate the gap filling to the presence of vortices in the system.

\item Since we used the mean-field dependence $\rho(T)$, it is clear
that the distance between the quasiparticle peaks (which is
approximately equal to $2 \rho$) diminishes as $T$ increases. This
process of the pseudogap closing is accompanied by the destruction
of the quasiparticle peaks. It is evident also that for $\rho = 0$,
the normal Fermi liquid behavior is immediately restored because
$J(\rho = 0) =0$ \cite{Gusynin.JETPLett,Gusynin.JETP}. Recall,
however, that the description proposed here cannot be applied when
$\rho$ is rather small, because, as already mentioned, the
fluctuations of $\rho(x)$ have to be also taken into account in
this region.

\end{enumerate}

\section{The density of states}
\setcounter{equation}{0}
\label{sec:DOS}

The density of states can be found from the formula
\begin{equation}
N(\omega) = \int \frac{d^2 k}{(2 \pi)^2} A(\omega, \mbox{\bf k}) =
N_0\int_{0}^{W} d \frac{k^2}{2m} A(\omega, \mbox{\bf k}),
\label{DOS.definition}
\end{equation}
where $N_0\equiv m/2 \pi$ is the density of 2D states in the normal
state ($W$ is the bandwidth).

This integral can be calculated analytically (see Appendix~\ref{sec:C}),
and which gives
\begin{eqnarray}
&& N(\omega)  = N_0\frac{\Gamma(\alpha)}{\Gamma(2-\alpha)} \left(
\frac{2 }{m r_{0}^{2}} \right)^{\alpha-1} \mbox{sgn} \omega\,
\theta(\omega^2 - \rho^2) \times \nonumber\\
&& \left\{({\cal A}_1)_{1 1} \left[\left(\frac{1}{2m \xi_{+}^{2}} +
W - \mu -\sqrt{\omega^{2} - \rho^{2}} \right)^{1-\alpha} -
\left(\frac{1}{2m \xi_{+}^{2}} \right)^{1-\alpha} \right]
\theta(\mu + \sqrt{\omega^{2} - \rho^{2}}) \right.
\nonumber\\
&& - \left. (\sqrt{\omega^{2} - \rho^{2}} \to -\sqrt{\omega^{2}
- \rho^{2}}) \right\}.
\label{DOS.final}
\end{eqnarray}

Again for $T =0$  and large $\mu \gg \rho$, Eq.~(\ref{DOS.final})
reduces to the BCS result \cite{Schrieffer}
\begin{equation}
N(\omega) = N_0\frac{|\omega|}{\sqrt{\omega^{2} - \rho^{2}}}.
\label{DOS.BCS}
\end{equation}

The plots for DOS (\ref{DOS.final}) are presented in Fig.~4 ($T <
T_{\rm BKT}$) and Figs.~5,6 for $T > T_{\rm BKT}$, respectively. If
one does not pay attention to a small difference in the curves shown
in these figures, it is well seen that qualitatively the form
of DOS does not differ from standard BCS curves. Moreover,
similarly to the spectral function, the DOS in the static
approximation has a gap both above and below $T_{\rm BKT}$ and does
not reveal any change when the temperature crosses the phase
transition point. This confirms once more the crossover character of
the latter, although, as was already pointed out, a 2D system is
superconducting below $T_{\rm BKT}$ only. According to generally
accepted views, the existence of an empty gap above the critical
temperature is impossible. The reasons for its persistence were
discussed in the previous section. Recall only that dynamical
fluctuations or fluctuations of the modulus $\rho$ undoubtedly
result in the gap filling above $T_{\rm BKT}$. One must also take
the dependence of the decay constant $\gamma$ into account (see
(\ref{correlator.dynamic})), which for $T>T_{\rm BKT}$ can be
essentially bigger than in the region $T<T_{\rm BKT}$
due to the presence of the vortices.

From the physical point of view, the filling of a gap
(transforming it into a pseudogap) above $T_{\rm BKT}$ (or $T_c$ in
quasi-2D case) has to continue up to $T^\ast$ (or $T_\rho$ if there
is a point where $\rho=0$). However, taking $\rho$-fluctuations
into account (i.e., $\rho(x)\rightarrow \rho+\Delta\rho$)
will cause the appearance of the self-energy, in addition to
$\rho^2$, in the denominator of the mean field Green's function
(\ref{Green.neutral}); it is proportional to the quantity
$\langle\Delta\rho(x)\Delta\rho(0)\rangle$ whose contribution could
persist at all $T$. In this case, the beginning of the pseudogap
opening will be defined by the experimental technique sensitivity
of the spectral function or DOS measurement.

\section{Conclusion}

To summarize, we have derived analytic expressions for the
fermion Green's function, its spectral density, and the density of
states in the modulus-phase representation for the simplest 2D
attractive Hubbard model with the $s$-wave nonretarded attractive
interaction.

While there is still no generally accepted microscopic theory of
HTSC compounds and their basic features (including the pairing
mechanism), it seems that this approach, although in a sense
phenomenological, is of great interest since it enables one to
propose a reasonable interpretation for the pseudogap phenomena
related to the vortex fluctuations. The results presented
here are entirely analytic, which allows a deeper understanding
than in the case of a numerical investigation. In particular, the
analytic investigation of the Green's function structure revealed
that the phase fluctuations lead to a non-Fermi liquid behavior below
and above $T_{\rm BKT}$.

Evidently, there are a number of important open questions.
The main question is whether the pseudogap is related to
some kind of superconducting (in our case, phase) fluctuations.
Hopefully, the experiment proposed in \cite{Scalapino} may answer
this question.  It seems plausible from the theoretical point of
view that superconducting fluctuations should contribute to the
pseudogap (see, however, \cite{Gooding}). Nevertheless, one cannot
exclude the possibility that the superconducting contribution may
be neither the only nor the main contribution.

Another open question is which approach allows one to obtain the
pseudogap from the attractive Hubbard model. The schemes used in
\cite{Franz} and in our paper are very different from those of
\cite{Levin}. In particular, our approach allowed us to establish a
direct relationship between the superconducting fluctuations and
the non-Fermi liquid behavior in a very natural and transparent way.
Also, it relates the pseudogap to the ``soup'' of fluctuating
vortices (see also \cite{Franz,Dorsey}), while \cite{Levin}
emphasises the existence of metastable pairs above $T_{c}$. It
is possible that both these pictures capture some physics, but in
the different regions of the temperatures. When $T$ is high and
close to $T^{\ast}$, the value of $\rho$ is small, so that
$\rho$-fluctuations or metastable pairs dominate. Then as the
temperature approaches $T_{\rm BKT}$, the values of $\rho$ and the
phase stiffness $J$ are growing bigger, so that the vortex
excitations dominate and $\rho$-fluctuations become less important.
We stress once more that the vortex excitations cannot be
adequately described within $T$-matrix approximation \cite{Micnas}.

Recently the last part of this picture was supported
experimentally \cite{Corson} by the measurements of the screening
and dissipation of a high-frequency electromagnetic field in
bismuth-cuprate films. These measurements provide evidence for
a phase-fluctuation driven transition from the superconducting to
normal state.

Finally, there remains the problem of a more complete treatment of the
pseudogap in the modulus-phase variables. In particular, the
effects of dynamical phase fluctuations and the fluctuations of the
order field modulus must be considered. The latter are especially
important for the $d$-wave superconductor since
the modulus can be arbitrary small in the nodal directions.
In this case it will be important
again to check the complete structure of the Green's function,
especially its non-pole structure. Another important question
that has to be addressed is which factor is more important for
the gap filling, the spin-charge coupling proposed in
\cite{Franz} or dynamical phase fluctuations which, as
was shown in \cite{Capezzali}, also result in filling.

\section*{Acknowledgments}
We thank Prof. R.M.~Quick for many thoughtful comments on this
manuscript. One of us (S.G.S.) is  grateful to the members of the
Department of Physics of the University of Pretoria, especially
Prof. R.M.~Quick and Dr. N.J.~Davidson, for very useful points
and hospitality. S.G.S. thanks Prof. A.~Sudb{\o} for a useful
discussion.
S.G.S. also acknowledges the financial support
of the Foundation for Research Development, Pretoria.

\renewcommand{\theequation}{\Alph{section}.\arabic{equation}}
\appendix

\section{The asymptotic of the phase correlator}
\setcounter{equation}{0}
\label{sec:A}

To calculate the integral in Eq.~(\ref{exponent}), we first write
it as
\begin{eqnarray}
&& I \equiv \int_0^\infty e^{-q/ \Lambda}[1-J_0(qr)]\coth{a q} =
\nonumber                 \\
&& {1\over a} \int_0^\infty dt e^{-t/\Lambda a}
\left( \coth t - {1\over t} \right)
\left[ 1 -J_0 \left({r\over a}t \right) \right] +
{1\over a} \int_0^\infty \frac{dt}{t} e^{-t/\Lambda}[1-J_0( rt)].
\label{A1}
\end{eqnarray}
The following formulas are used when calculating $I$:
\begin{eqnarray}
&& \int_0^\infty dte^{-\beta t}
\left( \coth t -{1\over t} \right) = \ln{\beta\over2}
+ {1\over\beta} - \psi \left(1 + \frac{\beta}{2} \right);
\nonumber\\
&& \int_0^\infty \frac{dt}{t} e^{-pt} [1-J_0(ct)] =
\ln\frac{p + \sqrt{p^2+c^2}}{2p}.
\label{A2}
\end{eqnarray}
Hence, we obtain
\begin{eqnarray}
I & = & {4\over r_0} \left[ \ln
\frac{1 + \sqrt{1+(\Lambda r)^2}}{\Lambda r_0} + {\Lambda r_0\over 4}-
\psi \left( 1 + {2\over\Lambda r_0} \right) \right] -
{1\over r}\int_0^\infty dt e^{-t/ \Lambda r}
\left( \coth \frac{r_0t}{4r} - \frac{4r}{r_0t} \right) J_0(t)
\nonumber                 \\
& \sim & {4\over r_0} \left[ \ln\frac{r}{r_0}+ {\Lambda r_0\over
4}-\psi \left( 1 + {2\over\Lambda r_0} \right) \right] - {1\over r}
\frac{1}{\sqrt{1 + 1/(\Lambda r)^2 }},
\qquad r \gg r_0, \Lambda^{-1}.
\label{A3}
\end{eqnarray}
Now, depending on the relationship between $\Lambda$ and $r_0$, we
obtain
\begin{equation}
I \sim
\left\{\begin{array}{cc}
{\frac{\ds 4}{\ds r_0}} \ln \frac{\ds r}{\ds R_0}+\Lambda, &  r \gg
r_0\gg \Lambda^{-1} \\
\frac{\ds 4}{\ds r_0} \ln \frac{\ds \Lambda r}{\ds 2},
&  r \gg \Lambda^{-1} \gg r_0,
\end{array} \right.
\label{A4}
\end{equation}
which gives Eq. (\ref{asymptotic}).

\section{Another representation for the retarded Green's function}
\setcounter{equation}{0}
\label{sec:B}

Here, we obtain another representation for the retarded
fermion Green's function that is more convenient for the derivation
of the spectral density. Recall that when the imaginary part of
$G(\omega,
\mbox{\bf k})$ is nonzero, $\mu +\sqrt{\omega^{2}
- \rho^{2}} > 0$ and $v_1 > 0$, $v_2 < 0$. This allows one to
transform the analytically continued (by means of Eq.
(\ref{roots.real})) integral
\begin{equation}
L \equiv \int_{0}^{\infty} du
\frac{[u (u + 1)]^{\alpha - 1}} {[(u + v_1)(u + v_2)]^{\alpha}}
\label{L}
\end{equation}
from Eq. (\ref{Green.final.Matsubara}) as
(for $\alpha<1$)
\begin{eqnarray}
& &L = (-1)^{\alpha} \int_{0}^{|v_2|} du
\frac{[u (u + 1)]^{\alpha - 1}} {[(u + v_1) ( |v_2| - u)]^{\alpha}} +
\int_{|v_2|}^{\infty} du
\frac{[u (u + 1)]^{\alpha - 1}} {[(u + v_1) (u - |v_2|)]^{\alpha}}
= \frac{(-1)^{\alpha}}{u_1^{\alpha}} \Gamma(\alpha)\times\nonumber\\
&&\Gamma(1-\alpha)
 F_1 \left(\alpha, \alpha, 1-\alpha; 1;
\frac{v_2}{v_1}, u_2 \right)+
\frac{1}{|v_2|}\frac{\Gamma(1-\alpha)}{\Gamma(2-\alpha)}
 F_1\left(1, \alpha, 1-\alpha; 2-\alpha;
\frac{v_1}{v_2}, \frac{1}{v_2} \right).
\label{L.transformation}
\end{eqnarray}
The first Appell function in (\ref{L.transformation}) can be
reduced to the hypergeometric function using the identity
\cite{Bateman} taht is valid for $\gamma = \beta + \beta^{\prime}$
\begin{equation}
F_1 (\alpha, \beta, \beta^{\prime}, \beta + \beta^{\prime}; x,y ) =
(1-y)^{-\alpha} {}_2F_{1} \left( \alpha, \beta;
\beta+\beta^{\prime}; \frac{x - y}{1 - y} \right).
\label{appelltohyper}
\end{equation}
Thus one obtains
\begin{eqnarray}
&&L = \frac{(-1)^{\alpha} \Gamma(\alpha)\Gamma(1-\alpha)} {[u_1
(1-u_2)]^{\alpha}}{}_2F_{1} \left(\alpha, \alpha;
1;\frac{u_2(1-u_1)} {u_1(1-u_2)} \right) + \frac{1}{|u_2|}
\frac{\Gamma(1-\alpha)}{\Gamma(2-\alpha)}\times\nonumber\\
&&F_1 \left(1, \alpha, 1-\alpha; 2-\alpha;\frac{u_1}{u_2},
\frac{1}{u_2} \right);\quad\frac{u_2(1-u_1)}{u_1(1-u_2)} < 1,
\quad \frac{u_1}{u_2} < 0,
\quad \frac{1}{u_2} < 0.
\label{L.final}
\end{eqnarray}
This completes the derivation of Eq. (\ref{Green.real.spectral}).

\section{The calculation of the density of states}
\setcounter{equation}{0}
\label{sec:C}

Introducing
\begin{equation}
y = \frac{{k^2/2m}}{\mu + \sqrt{\omega^{2} -
\rho^{2}}},
\qquad
b = \frac{1}{2m \xi_{+}^{2}} \frac{1}{\mu + \sqrt{\omega^{2} - \rho^{2}}}
\qquad y_0 = \frac{W}{\mu + \sqrt{\omega^{2} - \rho^{2}}}\,,
\label{C1}
\end{equation}
and substituting (\ref{spectral.density.final}) in
(\ref{DOS.definition}), we can write
\begin{eqnarray}
&& N(\omega) = N_0\frac{\Gamma(\alpha)}{\Gamma(1-\alpha)} \left(
\frac{2}{m r_{0}^{2}} \right)^{\alpha-1} \mbox{sgn} \omega\,
\theta(\omega^2 - \rho^2)
\left[ ({\cal A}_1)_{1 1} (\mu + \sqrt{\omega^{2} -
\rho^{2}})^{1-\alpha}\times\right.\nonumber\\
&&\left.\int_{0}^{y_0} \frac{dy}{[(y+b-1)^2 +4b]^{\alpha/2}}
{}_2F_1
\left( \frac{\alpha}{2}, \frac{1-\alpha}{2}; 1;
-\frac{4y}{(y+b-1)^2 +4b} \right)  \right.
\nonumber               \\
&& \left. \times \theta(\mu + \sqrt{\omega^{2} - \rho^{2}})
- (\sqrt{\omega^{2} - \rho^{2}} \to
- \sqrt{\omega^{2} - \rho^{2}}) \right].
\label{C2}
\end{eqnarray}

We now consider the integral from (\ref{C2}),
\begin{equation}
I = \int_{0}^{y_0} \frac{dy}{[(y+b-1)^2 + 4b]^{\alpha/2}}
{}_2F_1 \left( \frac{\alpha}{2}, \frac{1-\alpha}{2};1;
-\frac{4y}{(y+b-1)^2 + 4b} \right).
\label{C4}
\end{equation}
Using  the relation \cite{Bateman}
\begin{equation}
{}_2F_1(a, b; c; z) = (1-z)^{-a} {}_2F_1 \left(a, c-b; c;
\frac{z}{z-1} \right),
\label{C5}
\end{equation}
it can then be rewritten as
\begin{equation}
I  =  \int_{0}^{y_0} \frac{dy}{(y+b+1)^{\alpha}}
{}_2F_1 \left(\frac{\alpha}{2}, \frac{1+\alpha}{2}; 1;
\frac{4y}{(y+b+1)^2} \right).
\label{C6}
\end{equation}
Replacing $x = \frac{\ds b+1}{\ds y+b+1}$ in (\ref{C6}),
we obtain
\begin{equation}
I = (b+1)^{1-\alpha} \int_{x_0}^{1} dx x^{\alpha-2}
{}_2F_1 \left(\frac{\alpha}{2}, \frac{1+\alpha}{2}; 1;
\frac{4x(1-x)}{b+1} \right),\quad x_0 = \frac{b+1}{y_0+b+1}.
\label{C7}
\end{equation}

The integral (\ref{C6}) diverges at the lower limit as
$x_0 \to 0$ or equivalently as $y_0 \to \infty$.
To handle this we can write
\begin{eqnarray}
I & = & (b+1)^{1-\alpha} \int_{x_0}^{1} dx x^{\alpha-2}
\left[ {}_2F_1
\left(\frac{\alpha}{2}, \frac{1+\alpha}{2};1;
\frac{4x(1-x)}{b+1} \right) -1+1 \right]
\nonumber\\
& = & (b+1)^{1-\alpha} \left\{ \frac{1-x_0^{\alpha-1}} {\alpha-1} +
\int_{x_0}^{1} dx x^{\alpha-2} \left[
{}_2F_1 \left( \frac{\alpha}{2}, \frac{1+\alpha}{2}; 1;
\frac{4x(1-x)}{b+1} \right) -1 \right]\right\}.
\label{C9}
\end{eqnarray}
To calculate the last integral in (\ref{C9}), we rewrite it as
\begin{equation}
E = \lim_{\gamma \to \alpha-1} \int_{0}^{1} dx x^{\gamma-1}
\left[{}_2F_1 \left(\frac{\alpha}{2}, \frac{1+\alpha}{2};1;
\frac{4x(1-x)}{b+1} \right) -1\right].
\label{C10}
\end{equation}
For $\gamma>0$ we can compute the integral with the help of the formula
(2.21.29) \cite{Prudnikov.vol3}
\begin{eqnarray}
&& \int_{0}^{y} x^{\alpha-1} (y-x)^{\beta-1}
{}_2F_1(a,b;c;\omega x(y-x)) dx
\nonumber\\
&& = y^{\alpha+\beta-1} B(\alpha,\beta)
{}_4F_3 \left(a, b, \alpha, \beta; c,
\frac{\alpha+\beta}{2}, \frac{\alpha+\beta+1}{2};
\frac{\omega y^4}{4}\right),
\qquad y,\, \mbox{Re}\alpha,\, \mbox{Re} \beta > 0,
\label{C11}
\end{eqnarray}
so that
\begin{eqnarray}
E & = & \lim_{\gamma \to \alpha-1} \left\{ B(\gamma,1)
{}_4F_3 \left(\frac{\alpha}{2}, \frac{1+\alpha}{2}, \gamma, 1; 1,
\frac{\gamma+1}{2}, \frac{\gamma+2}{2}; \frac{1}{b+1} \right) -
\frac{1}{\gamma} \right\}
\nonumber\\
& = & B(\alpha-1,1) _{4}F_3 \left(\frac{\alpha}{2},
\frac{1+\alpha}{2}, \alpha-1, 1; 1, \frac{\alpha}{2},
\frac{1+\alpha}{2}; \frac{1}{b+1} \right) - \frac{1}{\alpha-1}
\nonumber\\
& = & \frac{1}{\alpha-1}{}_1F_0
\left(\alpha-1; \frac{1}{b+1} \right) - \frac{1}{\alpha-1}
 =  \frac{1}{1-\alpha} \left[ 1- \left(
\frac{b}{b+1}\right)^{1-\alpha} \right] > 0.
\label{C12}
\end{eqnarray}
Thus, for the integral (\ref{C4}) we find
\begin{equation}
I =\frac{1}{1-\alpha} [(y_0+b+1)^{1-\alpha} - b^{1-\alpha}].
\label{C14}
\end{equation}
Now substituting (\ref{C14}) into (\ref{C2}) we obtain
\begin{eqnarray}
& &N(\omega) = N_0\frac{\Gamma(\alpha)}{\Gamma(2-\alpha)} \left(
\frac{2}{m r_0^2} \right)^{\alpha-1} \mbox{sgn} \omega\,
\theta(\omega^2 - \rho^2)\left\{({\cal A}_1)_{1 1} (\mu + \sqrt{\omega^{2}
 -\rho^{2}})^{1-\alpha}\right.\nonumber\\
& &\left. \times\left [(y_0+b+1)^{1-\alpha} - b^{1-\alpha}\right]
\theta(\mu +\sqrt{\omega^{2} - \rho^{2}})
- (\sqrt{\omega^{2} - \rho^{2}} \to -\sqrt{\omega^{2} -
\rho^{2}})\right\}.
\label{C15}
\end{eqnarray}
Finally, replacing $y_0$ and $b$ in (\ref{C15}) by expressions from
(\ref{C1}) we arrive at Eq. (\ref{DOS.final}).

\newpage

\smallskip

\begin{figure}
\centerline{
\epsfxsize=6cm
\epsfbox[20 355 275 530]{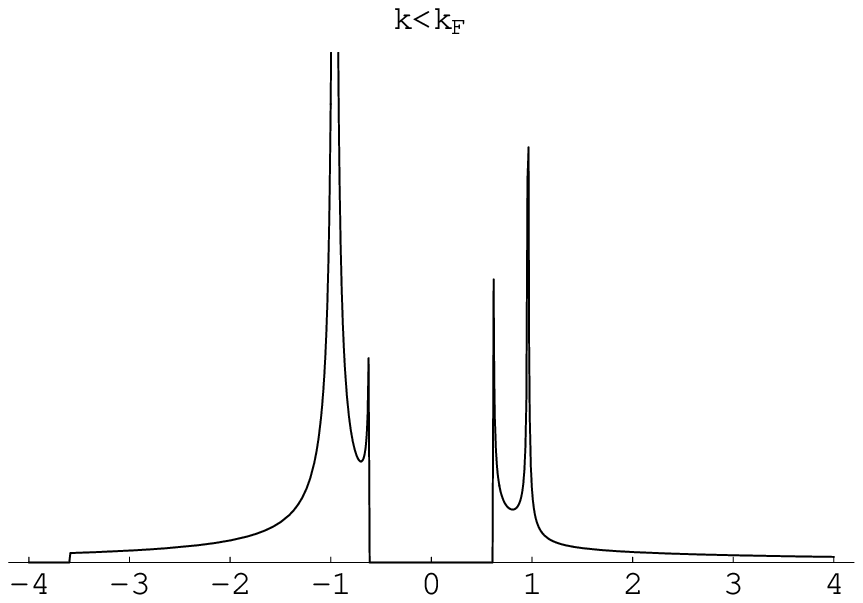}
\epsfxsize=6cm
\epsfbox[20 355 275 530]{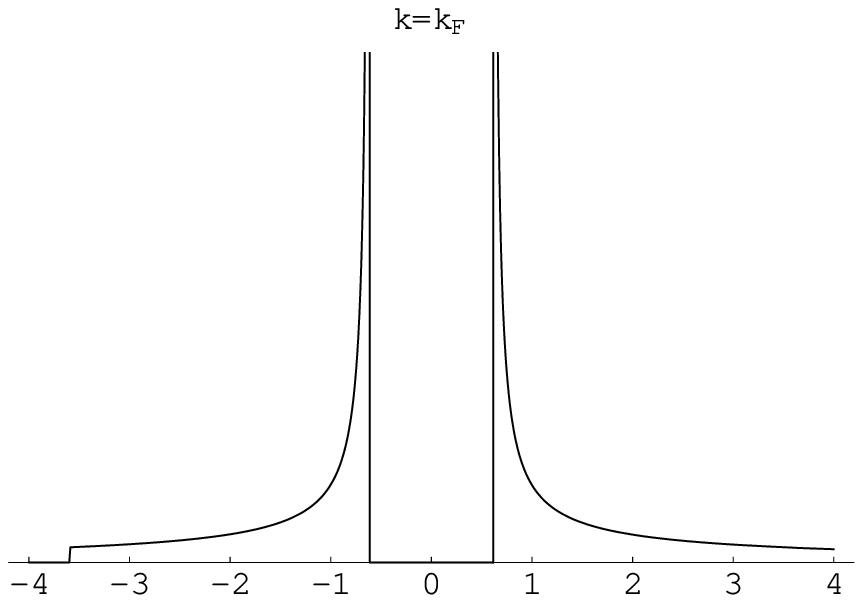}
\epsfxsize=6cm
\epsfbox[20 355 275 530]{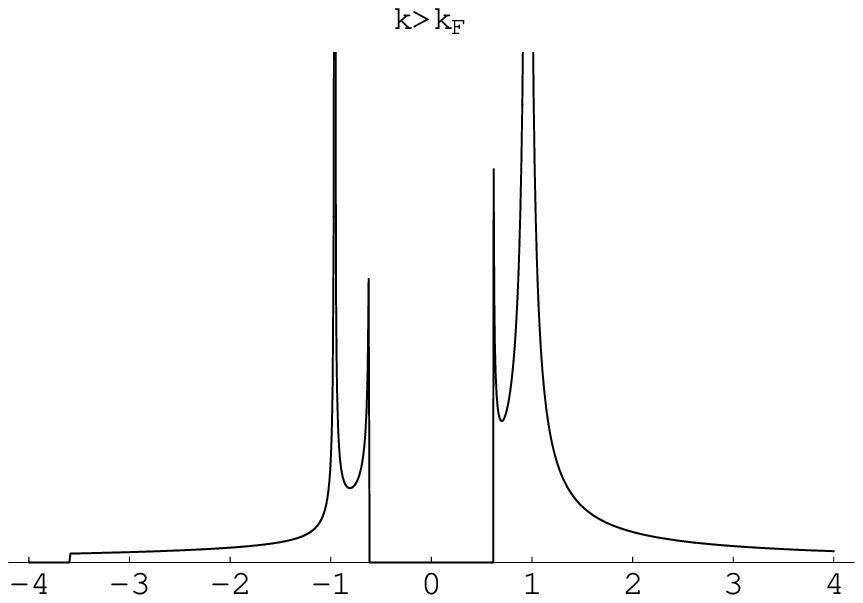}
}
\vspace{1cm}
\caption{
Plot of the spectral function $A(\omega , \mbox{\bf k}) $ as a
function of $\omega$ in units of the zero temperature gap $\Delta$
for $k < k_{F}$, $k = k_{F}$ and $k > k_{F}$ at $T = 0.99 T_{\rm
BKT}$.}
\label{fig1}
\end{figure}

\begin{figure}
\centerline{
\epsfxsize=6cm
\epsfbox[20 355 275 530]{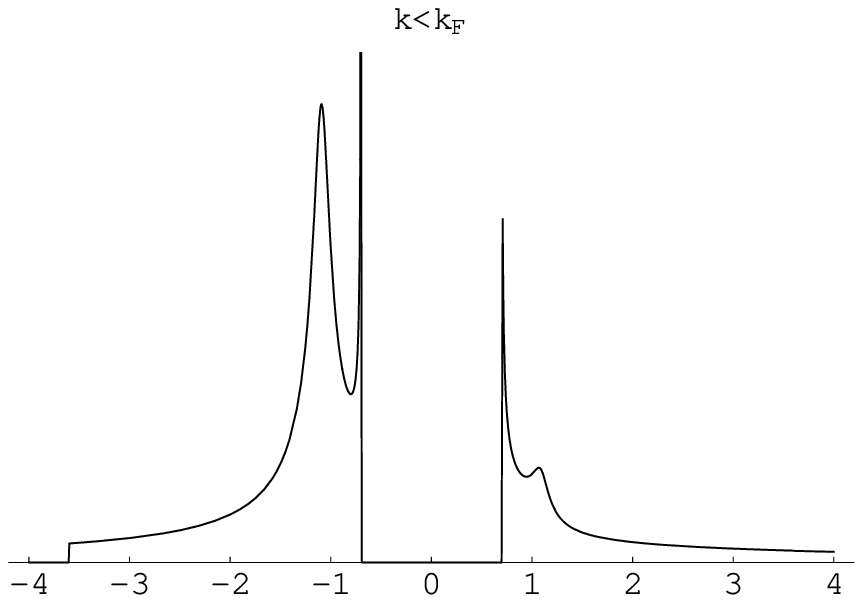}
\epsfxsize=6cm
\epsfbox[20 355 275 530]{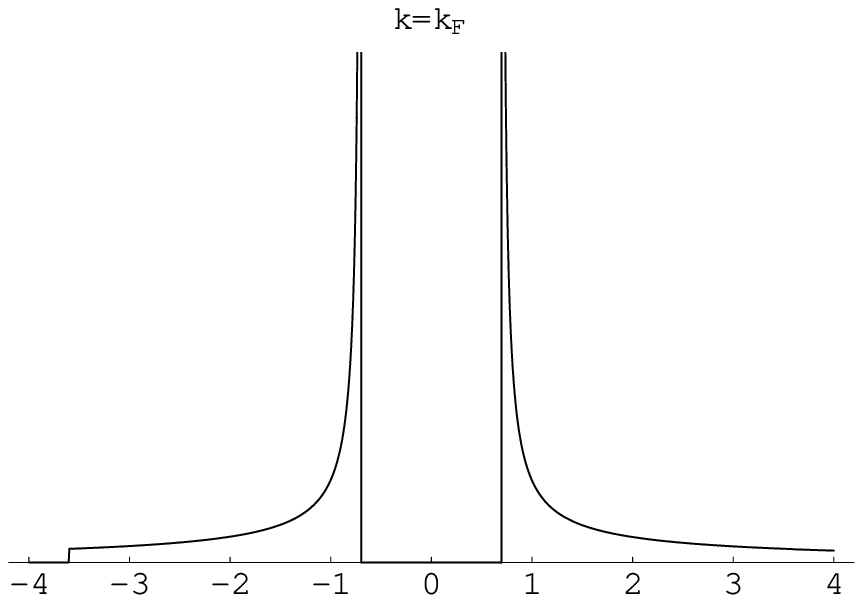}
\epsfxsize=6cm
\epsfbox[20 355 275 530]{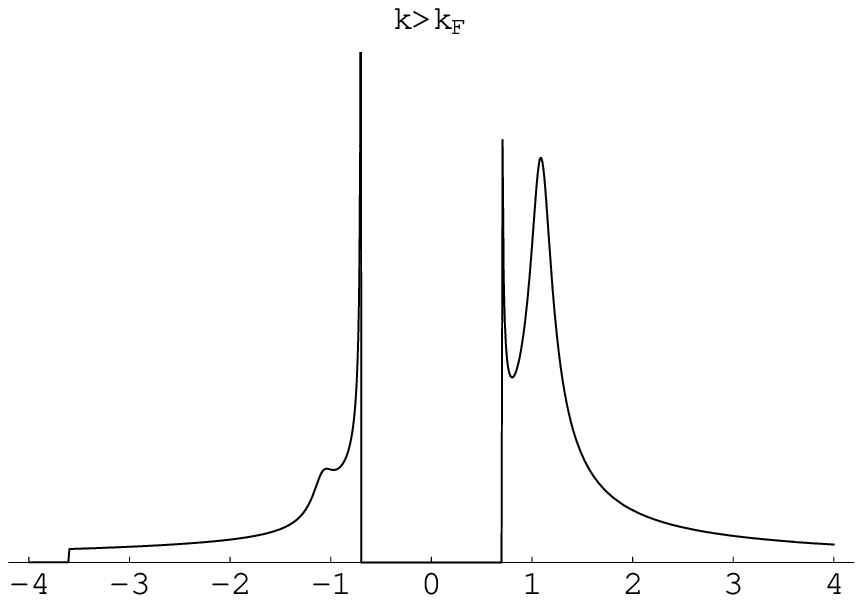}
}
\vspace{1cm}
\caption{
Plot of the spectral function $A(\omega , \mbox{\bf k}) $ as a
function of $\omega$ in units of the zero temperature gap $\Delta$
for $k < k_{F}$, $k = k_{F}$ and $k > k_{F}$ at $T = 1.043 T_{\rm
BKT}$.}
\label{fig2}
\end{figure}

\begin{figure}
\centerline{
\epsfxsize=6cm
\epsfbox[20 355 275 530]{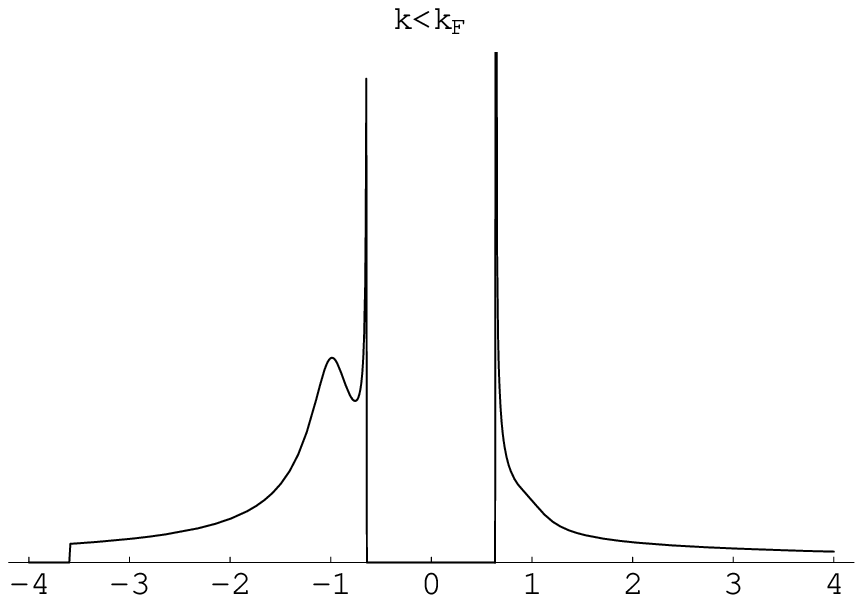}
\epsfxsize=6cm
\epsfbox[20 355 275 530]{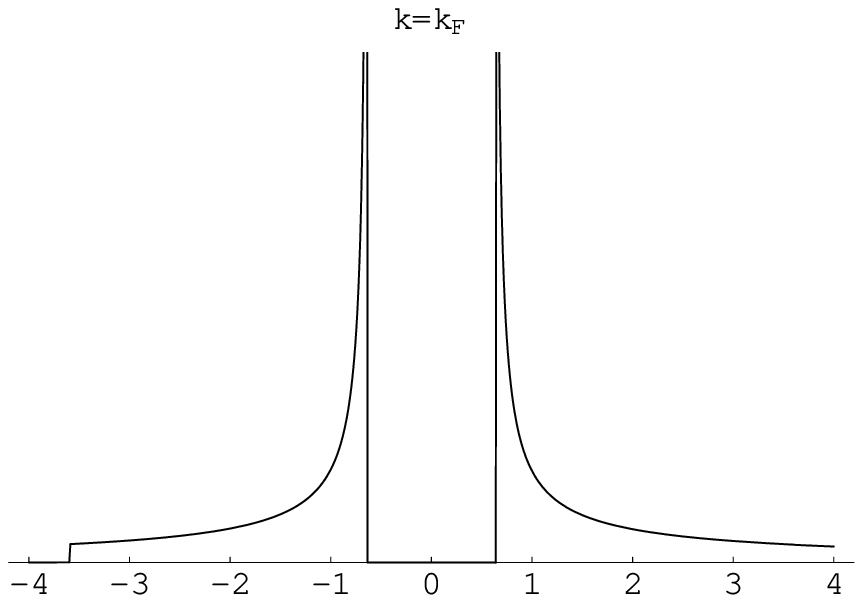}
\epsfxsize=6cm
\epsfbox[20 355 275 530]{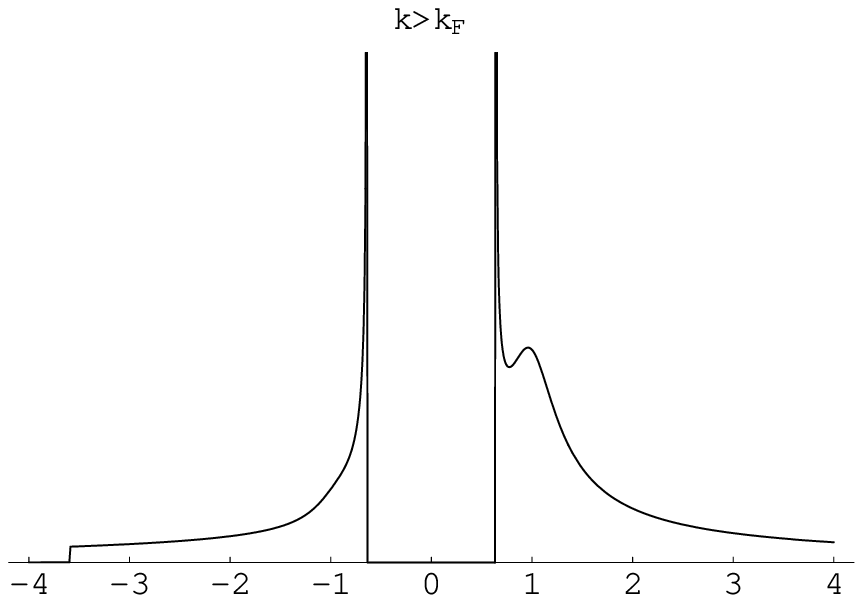}
}
\vspace{1cm}
\caption{
Plot of the spectral function $A(\omega , \mbox{\bf k}) $ as a
function of $\omega$ in units of the zero temperature gap $\Delta$
for $k < k_{F}$, $k = k_{F}$ and $k > k_{F}$ at $T = 1.088 T_{\rm
BKT}$.}
\label{fig3}
\end{figure}

\newpage

\begin{figure}
\centerline{
\epsfxsize=4.0in
\epsfbox[20 355 275 530]{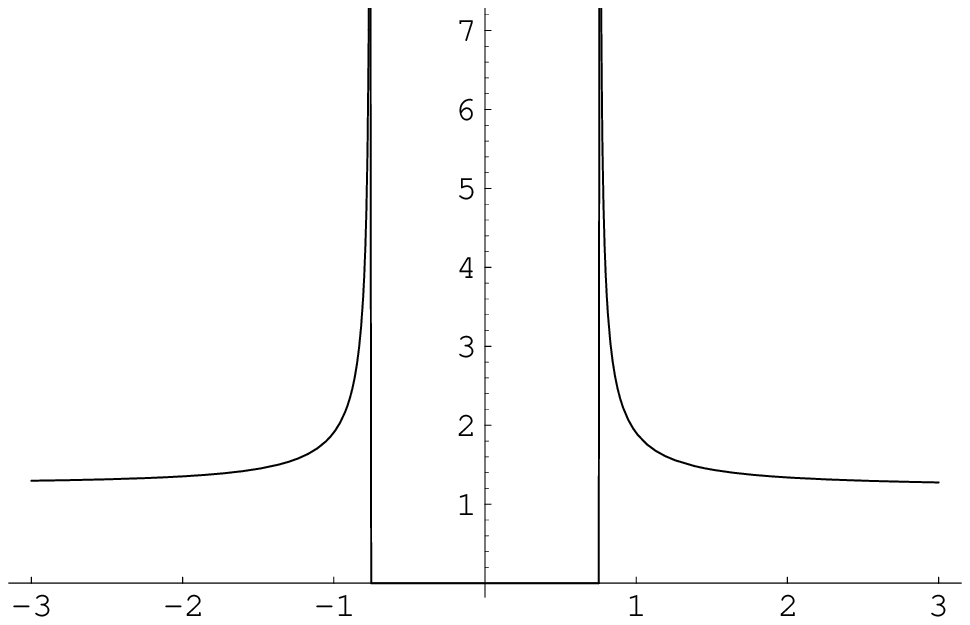}
}
\vspace{1cm}
\caption{The density of states $N(\omega)/N_0$  at
$T = 0.99 T_{\rm BKT}$.}
\label{fig4}
\end{figure}

\begin{figure}
\centerline{
\epsfxsize=4.0in
\epsfbox[20 355 275 530]{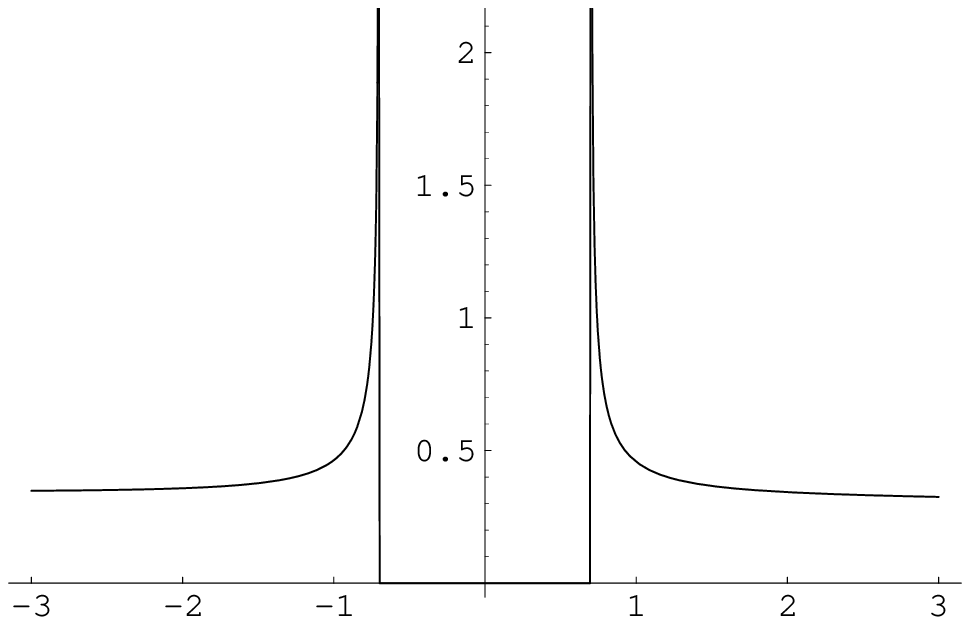}
}
\vspace{1cm}
\caption{The density of states $N(\omega)/N_0$  at
$T = 1.043 T_{\rm BKT}$.}
\label{fig5}
\end{figure}

\begin{figure}
\centerline{
\epsfxsize=4.0in
\epsfbox[20 355 275 530]{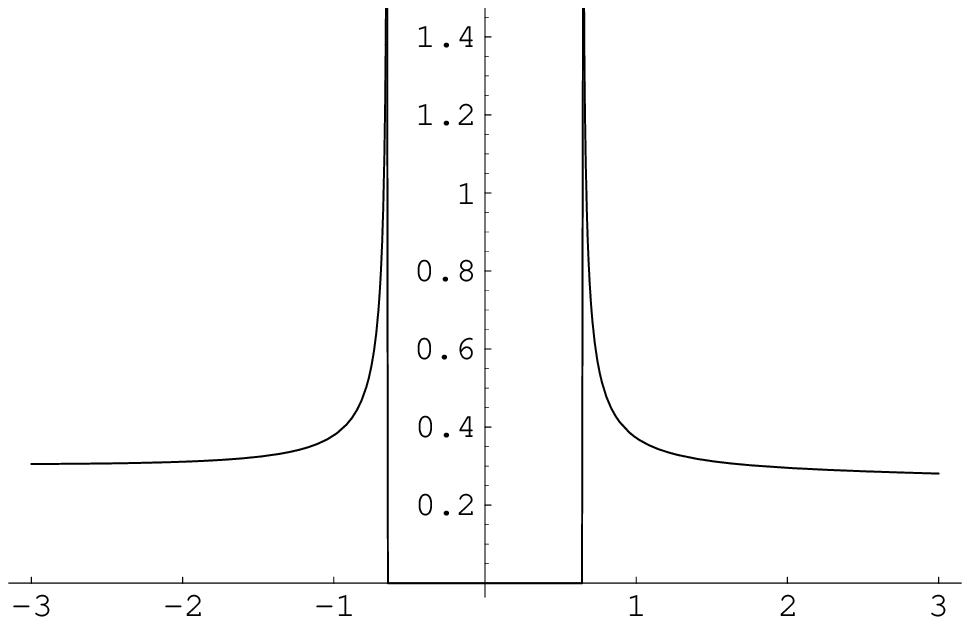}
}
\vspace{1cm}
\caption{The density of states $N(\omega)/N_0$  at
$T = 1.088 T_{\rm BKT}$.}
\label{fig6}
\end{figure}


\begin{thebibliography}{99}
\bibitem{foot} This in its turn means that the  approximation used
proves to be justified not far from optimal doping.
\end{thebibliography}
\end{document}